\documentclass[12pt]{article}
\usepackage{latexsym}
\newcommand{\eqn}[1]{(\ref{#1})}
\newcommand{\ft}[2]{{\textstyle\frac{#1}{#2}}}
\newsavebox{\uuunit}
\sbox{\uuunit}
    {\setlength{\unitlength}{0.825em}
     \begin{picture}(0.6,0.7)
        \thinlines
        \put(0,0){\line(1,0){0.5}}
        \put(0.15,0){\line(0,1){0.7}}
        \put(0.35,0){\line(0,1){0.8}}
       \multiput(0.3,0.8)(-0.04,-0.02){12}{\rule{0.5pt}{0.5pt}}
     \end {picture}}
\newcommand {\Cbar}
    {\mathord{\setlength{\unitlength}{1em}
     \begin{picture}(0.6,0.7)(-0.1,0)
        \put(-0.1,0){\rm C}
        \thicklines
        \put(0.2,0.05){\line(0,1){0.55}}
     \end {picture}}}
\newcommand {\unity}{\mathord{\!\usebox{\uuunit}}}
\newcommand  {\Rbar} {{\mbox{\rm$\mbox{I}\!\mbox{R}$}}}
\newcommand  {\Hbar} {{\mbox{\rm$\mbox{I}\!\mbox{H}$}}}

\def\mod{{\rm mod ~}}
\font\cmss=cmss10 \font\cmsss=cmss10 at 7pt
\def\ZZ{\relax\ifmmode\mathchoice
{\hbox{\cmss Z\kern-.4em Z}}{\hbox{\cmss Z\kern-.4em Z}}
{\lower.9pt\hbox{\cmsss Z\kern-.4em Z}}
{\lower1.2pt\hbox{\cmsss Z\kern-.4em Z}}\else{\cmss Z\kern-.4em
Z}\fi}

\csname @addtoreset\endcsname{equation}{section}

\newcommand{\Poin}{Poincar\'{e}}
\def\diag{{\rm diag}}
\newcommand{\dsl}{\not\!\partial}
\newcommand{\dr}{\raise.3ex\hbox{$\stackrel{\leftarrow}{\partial }$}{}}
\newcommand{\delr}{\raise.3ex\hbox{$\stackrel{\leftarrow}{\delta }$}{}}
\newcommand{\bbox}{\lower.2ex\hbox{$\Box$}}
\newcommand{\rmi}{{\rm i}}
\newcommand{\QED}{{\hspace*{\fill}\rule{2mm}{2mm}\linebreak}}
\def\trace{{\rm Tr}\hskip 1pt}

\begin{document}
\begin{titlepage}
\begin{flushright} October 1999\\
KUL-TF-99/37\\
hep-th/9910030
\end{flushright}
\vfill
\begin{center}
{\LARGE\bf Tools for supersymmetry}  $^1$   \\
\vskip 27.mm  \large
{\bf   Antoine Van Proeyen}$^2$ \\
\vskip 1cm
{\em Instituut voor theoretische fysica}\\
{\em Universiteit Leuven, B-3001 Leuven, Belgium}
\end{center}
\vfill

\begin{center}
{\bf ABSTRACT}
\end{center}
\begin{quote}
This is an elementary introduction to basic tools of supersymmetry:
the spacetime symmetries, gauge theory and its application in
gravity, spinors and superalgebras.
Special attention is devoted to conformal and anti-de Sitter
algebras.
\vfill      \hrule width 5.cm
\vskip 2.mm
{\small
\noindent $^1$ Lectures in the spring school
\emph{Q.F.T., Supersymmetry and Superstrings}, in C\v{a}lim\v{a}ne\c{s}ti,
Romania, April 1998.
\\
\noindent $^2$ Onderzoeksdirecteur FWO, Belgium;\\ E-mail:
Antoine.VanProeyen@fys.kuleuven.ac.be}
\end{quote}
\end{titlepage}
\tableofcontents{}
\newpage
\section{Introduction} \label{s:intro}
In this review you will not find many general ideas. It is rather
meant as a technical introduction to some basic tools for working
with supersymmetric field theories. It are ingredients which I could
 use for various applications in the last 20 years.

In section~\ref{s:bosstsym}, I will consider only bosonic theories. It
contains an introduction to the spacetime symmetries. Also gauging of
symmetries is treated, and special attention is devoted to a very
useful theorem for calculating the transformation of the covariant
derivative of fields.

Section~\ref{ss:spinors} reviews the properties of
Clifford algebras and spinors in a spacetime with arbitrary
dimensions and signature. It gives an overview of Majorana, Weyl,
Majorana--Weyl and symplectic Majorana--Weyl spinors. At the end
several tips are given to perform complex conjugation in
practice, and to manipulate gamma matrices.

Supersymmetry is introduced in section~\ref{s:susyalg}. I introduce
especially the algebraic approach.
Also the elementary facts on superalgebras are treated, including the
real forms of these algebras. The aim is to come to supersymmetry
algebras, as well super-\Poin, as super-adS and superconformal
algebras.

\section{Bosonic spacetime symmetries}\label{s:bosstsym}
Supersymmetry is related to the structure of spacetime.
Before tackling supersymmetry algebras, I will first review the
bosonic spacetime symmetries and their gauging.
\subsection{Coleman--Mandula result}\label{ss:ColemanMandula}
To investigate possible spacetime symmetries, Coleman and Mandula
\cite{ColemanMandula} investigated how large the spacetime symmetry
group can be in order that scattering amplitudes do not become
trivial. They thus deal with `visible symmetries', i.e.\ those which
act on $S$-matrix elements. They restrict themselves also to 4
dimensions, and assume a finite number of different particles in a
multiplet. Their research was before the advent of supersymmetry,
so they were looking only at bosonic symmetries. Still the result is
very significant. There are two cases to be distinguished.

\noindent 1. There are massive particles. Then the symmetry algebra
has at most translations $P_\mu$, Lorentz rotations $M_{\mu\nu}$
and an algebra $G$ of scalar (i.e.\ commuting with $P$ and $M$)
symmetries.

\noindent 2. There are only massless particles. Then there is the
extra possibility of having conformal symmetry (and a commuting group
$G$).

The second possibility will be discussed in section~\ref{ss:rigidconf}.
The first one contains a direct sum of $G$ with the Poincar\'{e}
algebra, i.e.\
\begin{eqnarray}
[M_{\mu\nu} , M^{\rho\sigma}]&=&
-2\delta_{[\mu}^{[\rho} M_{\nu]}{}^{\sigma]} \,, \nonumber\\
\ [P_{\mu} , M_{\nu\rho}\ ] &=&\eta_{\mu[\nu} P_{\rho]}\,, \nonumber\\
\ [P_{\mu} , P_{\nu}\ ] &=&  0 \,.
\label{Poinalgebra}
\end{eqnarray}
For conventions, see appendix~\ref{app:conv}. In supersymmetry
 the last commutator is often modified, giving rise to
the (anti) de Sitter algebra.

\subsection{Anti-de Sitter algebra and spacetime}\label{ss:adS}
The Poincar\'{e} algebra is an `In\"{o}n\"{u}--Wigner' contraction of the
`(anti) de Sitter algebra':
\begin{eqnarray}
\left[M_{\mu\nu},M_{\rho\sigma}\right]
&=&\eta_{\mu[\rho}M_{\sigma]\nu}
-\eta_{\nu[\rho}M_{\sigma]\mu}\nonumber\\
\left[P_\mu , M_{\nu\rho}\right]&=&\eta_{\mu[\nu}P_{\rho]}\nonumber\\
\left[ P_\mu,P_\nu\right]&=& \frac{1}{2 R^2}M_{\mu\nu}\,.
\label{adSalgebra}
\end{eqnarray}
With the opposite sign for the last commutator we would have the
`de Sitter algebra'.
As written it is the `anti-de Sitter ($adS$) algebra'. And for
$R\rightarrow \infty$, we recover the \Poin\ algebra. Defining
 $M_{d\mu}=-M_{\mu d}= R\, P_\mu$ we have generators
$M_{\hat\mu\hat \nu}=-M_{\hat\nu\hat \mu}$ with $\hat\mu=0,\ldots
,d$, and defining the metric to be $\eta_{\hat\mu\hat \nu}=\diag
(-+\ldots +-)$ the algebra can be concisely written as
\begin{equation}
\left[M_{\hat\mu\hat\nu},M_{\hat\rho\hat\sigma}\right]
=\eta_{\hat\mu[\hat\rho}M_{\hat\sigma]\hat\nu}
-\eta_{\hat\nu[\hat\rho}M_{\hat\sigma]\hat\mu} \,,
\end{equation}
i.e.\ it is the algebra $SO(d-1,2)$. Note that every point is
invariant under the rotations around this point, while not invariant
under the translations. In this sense we can write
\begin{equation}
adS_d= \frac{SO(d-1,2)}{SO(d-1,1)}\,.
\end{equation}
The `In\"{o}n\"{u}---Wigner' contraction  is the statement that the $SO(d-1,2)$
algebra reduces to the \Poin\ algebra when taking the limit $R\rightarrow
\infty$ keeping $P_\mu$ constant in the relation $M_{d\mu}=-M_{\mu d}= R\, P_\mu$.

To obtain a space with $adS$ metric, we start from defining it as
a submanifold of a $(d+1)$-dimensional space with a flat metric of
$(d-1,2)$ signature
(for convenience we taken here $\mu=0,\ldots ,d-2$)
\begin{equation} \tabcolsep 1pt
\begin{array} {ccccc}
ds^2=&dX^\mu \eta_{\mu\nu} dX^\nu&-&dX^+dX^- &      \\
     & (d-2,1)                     &+&(1,1)    &\Rightarrow (d-1,2)
\end{array}\tabcolsep 6pt
\label{metricind+1}
\end{equation}
The $adS$ space is the submanifold determined by the
$SO(d-1,2)$-invariant equation
\begin{equation}
X^\mu \eta_{\mu\nu} X^\nu-X^+X^- +R^2 =0\,.  \label{adSsubmanif}
\end{equation}
On the hypersurface one can take several sets of coordinates. E.g.\ the
horospherical coordinates $\{x^\mu,z \}$ are defined by
\begin{eqnarray}
X^-&=&z^{-1}\nonumber\\
X^\mu &=& z^{-1}  x^\mu\nonumber\\
X^+&=&\frac{x_\mu^2+R^2  z^2 }{z }  \,.
\end{eqnarray}
The latter being the solution of \eqn{adSsubmanif} given the first two.
The induced metric on the hypersurface is
\begin{equation}
ds^2=\frac{1}{z^2}\left(  dx_\mu^2+R^2 dz ^2 \right) \,. \label{adSmetric}
\end{equation}
The $SO(d-1,2)$ is linearly realized in the embedding $(d+1)$-dimensional
space, and these transformations, ($\hat\mu=\mu,+,-$ and
$\Lambda^{\hat\mu\hat\nu}=-\Lambda^{\hat\nu\hat\mu}$)
\begin{equation}
\delta X^{\hat\mu}=\Lambda^{\hat\nu\hat\rho}M_{\hat\nu\hat\rho}
X^{\hat\mu}=-\Lambda^{\hat\mu}{}_{\hat\nu} X^{\hat\nu}\,.
\end{equation}
are on the $adS$ space distorted to\footnote{Note that to raise or
lower indices one has to use the metric in (\ref{metricind+1}), i.e.\
$\eta _{+-}=-\ft12$, and thus $\eta ^{+-}=-2$.}
\begin{eqnarray}
\delta_{adS} x^\mu&=& -\Lambda^\mu{}_- -\Lambda^{\mu\nu}x_\nu
-\Lambda^+{}_+ x^\mu \nonumber\\ &&
-(x^2+R^2 z^2)\Lambda^\mu{}_++2x^\mu x_\nu \Lambda^\nu{}_+  \nonumber\\
\delta_{adS} z&=&-z\left( \Lambda^+{}_+ -2x_\mu\Lambda^\mu{}_+
\right)\,.   \label{adSiso}
\end{eqnarray}

Suppose I just gave you the metric \eqn{adSmetric}. Then to determine the
symmetries, one should solve the `Killing' equation (with $X^M=\{x^\mu,\phi\}$)
\begin{equation}
(\partial_P g_{MN}) \delta X^P+2 g_{P(M}\partial_{N)}\delta X^P =0\,,
\end{equation}
which is the requirement that the Lie derivative of the metric vanishes.
As an \textit{exercise}, one may prove that the transformations \eqn{adSiso} are the
only solutions.

\subsection{Rigid conformal symmetry}\label{ss:rigidconf}
Conformal symmetry is defined as the symmetry which preserves angles. Therefore
it should contain the transformations which change the metric up to a
factor. That implies that the symmetries are determined by the
solutions to the `conformal Killing equation'
\begin{equation}
\partial_{(\mu}\xi_{\nu)}-\ft1d \eta_{\mu\nu}\partial_\rho \xi^\rho=0\,.
\end{equation}
In $d=2$ with as non-zero metric elements $\eta_{z\bar z}=1$,
the  Killing equations are reduced to
$\partial_z  \xi _z = \partial_{\bar z}  \xi_{\bar z} = 0$
and this leads to an infinite dimensional conformal algebra (all holomorphic
vectors $ \xi_{\bar z}(z)$ and anti-holomorphic vectors $ \xi _z(\bar z) $).
In dimensions $d>2$ the conformal algebra is finite-dimensional. Indeed, the
solutions are
\begin{equation}
\xi^\mu(x)=a^\mu +\lambda_M^{\mu\nu}x_\nu+\lambda_D x^\mu
+(x^2\Lambda_K^\mu-2x^\mu x\cdot \Lambda_K) \,. \label{ximu}
\end{equation}
Corresponding to the parameters $a^\mu$ are the translations $P_\mu$,
to $\lambda_M^{\mu\nu}$ correspond the Lorentz rotations $M_{\mu\nu}$, to
$\lambda_D$ are associated dilatations $D$, and $\Lambda_K^\mu$ are parameters
of `special conformal transformations' $K_\mu$. This is expressed as follows
for the full set of conformal transformations $\delta_C$:
\begin{equation}
\delta_C= a^\mu  P_\mu + \lambda_M^{\mu\nu}M_{\mu\nu}+\lambda_D D +
\Lambda_K^\mu K_\mu   \,.
\end{equation}
With these transformations, one can obtain the algebra with as non-zero commutators
\begin{eqnarray}
&&[M_{\mu\nu} , M^{\rho\sigma}]=
-2\delta_{[\mu}^{[\rho} M_{\nu]}{}^{\sigma]} \,, \nonumber\\
&& [P_{\mu} , M_{\nu\rho}\ ] =\eta_{\mu[\nu} P_{\rho]}\,, \qquad 
 [K_{\mu} , M_{\nu\rho}\ ] = \eta_{\mu[\nu}K_{\rho]} \,, \nonumber\\
&& [P_{\mu} , K_{\nu}\ ]= 2 (\eta_{\mu\nu} D + 2 M_{\mu\nu}) \,,
\nonumber\\
&& [D , P_{\mu} \ ]= P_\mu \,,\qquad \quad \quad
\ [D , K_{\mu} \ ]=-K_\mu \,.     \label{confalg}
\end{eqnarray}
This is the $SO(d,2)$ algebra\footnote{In the
2-dimensional case $SO(2,2)=SU(1,1)\times SU(1,1)$ is realised by
the finite subgroup of the infinite dimensional
conformal group, and is well known in
terms of $L_{-1}= {1\over 2} (P_0 - P_1)$, $L_0={1\over 2}D+M_{10}$,
$L_1={1\over 2} (K_0 + K_1)$,
$\bar L_{-1}= {1\over 2} (P_0 + P_1)$, $\bar L_0={1\over 2}D-M_{10}$,
$\bar L_1={1\over 2} (K_0 - K_1)$. Higher order $L_n, |n| \geq 2 $ have
no analogs in $d > 2$. }. Indeed one can define
\begin{equation}
M^{\hat \mu\hat \nu}=\pmatrix{M^{\mu\nu}&\ft14(P^\mu-K^\mu)&
\ft14(P^\mu+K^\mu)\cr -\ft14(P^\nu-K^\nu)&0&-\ft12 D\cr
-\ft14(P^\nu+K^\nu) &\ft12 D &0  }\,,
\end{equation}
where indices are raised w.r.t. the rotation matrices $M^\mu {}_\nu $
with the metric $\eta={\rm diag~}(-1,1,...,1,-1)$.  Note that this is
the same as the anti-de Sitter algebra in $d+1$ dimensions
\begin{equation}
Conf_{d} = adS_{d+1}\,,\label{confisadS}
\end{equation}
(obviously this concerns the algebras, not the spaces)
which is an essential ingredient in the adS/CFT correspondence which got
recently so much attention. Note in this respect the similarity between
\eqn{adSiso} and \eqn{ximu}.

In general, fields $\phi^i(x)$ in $d$ dimensions have the following
transformations under the conformal group:
\begin{eqnarray}
\delta_C \phi^i(x)&=& \xi^\mu(x)\partial_\mu \phi^i(x)
+ \Lambda_M^{\mu\nu}(x)\, m_{\mu\nu}{}^i{}_{j}\phi^j(x)
\nonumber\\ &&+ w_i\,\Lambda_D(x)
\,\phi^i(x)+ \Lambda_K^\mu\left( k_\mu \phi\right) ^i(x)\,,
\label{deltaC}
\end{eqnarray}
where the
$x$-dependent rotation $\Lambda_{M\,\mu\nu}(x)$ and $x$-dependent
dilatation $\Lambda_D(x)$ are given by
\begin{eqnarray}
\Lambda_{M\,\mu\nu}(x)&=&\partial_{[\nu}\xi_{\mu]}=\lambda_{M\,\mu\nu}
-4x_{[\mu} \Lambda_{K\,\nu]} \,, \nonumber\\
\Lambda_D(x)&=&\ft1d \partial_\rho \xi^\rho =
\lambda_D -2 x\cdot \Lambda_K  \,.   \label{Lambdax}
\end{eqnarray}
To specify for each field $\phi^i$ its transformations under conformal
group one has to specify:
\begin{description}
\item[i) transformations under the Lorentz group,]
encoded into the matrix
$(m_{\mu\nu})^i{}_j$. The Lorentz transformation matrix $m_{\mu\nu}$ should
satisfy
\begin{equation}
m_{\mu\nu}{}^i{}_k m_{\rho\sigma}{}^k{}_j  -
m_{\rho\sigma}{}^i{}_k m_{\mu\nu}{}^k{}_j =
-\eta_{\mu[\rho}m_{\sigma]\nu}{} ^i{}_j
+\eta_{\nu[\rho}m_{\sigma]\mu} {}^i{}_j \,. \label{algmatrlor}
\end{equation}
The explicit form for Lorentz transformation matrices is for
vectors (the indices $i$ and $j$ are of the same kind as $\mu$
and $\nu$)
\begin{equation}
m_{\mu\nu}{}^\rho{}_\sigma =-\delta^\rho_{[\mu}\eta_{\nu]\sigma} \,,
\end{equation}
while for spinors, (where $i$ and $j$ are (unwritten) spinor indices)
\begin{equation}
m_{\mu\nu} =-\ft14\gamma_{\mu\nu}\,.
\end{equation}
\item[ ii) The Weyl weights] $w_i$.
\item[ iii) Possible extra parts of the
special conformal transformations,] apart from those
connected to translations, rotations and dilatations as
in \eqn{ximu} and \eqn{Lambdax},  $(k_\mu \phi)^i$.
The Weyl weight of $(k_\mu \phi^i)$ should be
$w-1$, and $k_\mu$ are mutually commuting operators.
\end{description}
In this way, the algebra \eqn{confalg} is realised on the fields as
\begin{equation}
\left[\delta_C(\xi_1),\delta_C(\xi_2)\right]=\delta_C\left(
\xi^\mu=\xi_2^\nu\partial_\nu\xi_1^\mu-\xi_1^\nu\partial_\nu\xi_2^\mu\right)\,.
\end{equation}
Note the sign difference between the commutator of matrices \eqn{algmatrlor} and
the commutator of the generators in \eqn{confalg}, which is due to the difference
between `active and passive' transformations. To understand fully the meaning
of the order of the transformations, consider in detail the
calculation of the commutator of transformations of fields.
 See e.g.\  for a field of zero Weyl weight, and notice how the
transformations act only on fields, not on explicit spacetime points $x^\mu$:
\begin{eqnarray}
\lambda_D a^\mu [D,P_\mu]\phi(x)&=& \left( \delta_D(\lambda_D)
\delta_P(a^\mu)-\delta_P(a^\mu) \delta_D(\lambda_D)\right) \phi(x)
\nonumber\\ &=&
\delta_D(\lambda_D)  a^\mu \partial_\mu \phi(x)   - \delta_P(a^\mu)
\lambda_D x^\mu \partial_\mu \phi(x) \nonumber\\ &=&
a^\mu \partial_\mu  \left( \lambda_D x^\nu \right)  \partial_\nu
\phi(x)
\nonumber\\
&=& a^\mu \lambda_D \partial_\mu  \phi(x)  = \lambda_D a^\mu
P_\mu\phi(x) \,.
\end{eqnarray}
It is important to notice that the derivative of a field of Weyl weight $w$ has
weight $w+1$. E.g.\ for a scalar of weight $w$ (and without
extra special conformal transformations) we obtain
\begin{eqnarray}
\delta_C  \partial_\mu\phi(x)&=&\xi^\nu(x)\partial_\nu\partial_\mu \phi(x)
+w\,\Lambda_D(x)\,\partial_\mu\phi(x)\nonumber\\ &&
-\Lambda_{M\mu}{}^\nu (x)\partial_\nu \phi(x) +\Lambda_D(x)
\partial_\mu \phi(x) -2w\,\Lambda_{K\mu}  \phi(x) \,.  \label{delCderphi}
\end{eqnarray}
\par
With these rules the conformal algebra is satisfied. The
question remains when an action is conformal invariant. We consider
local actions which can be written as $S=\int d^dx\,{\cal
L}(\phi^i(x),\partial_\mu\phi^i(x))$, i.e.\ with at most first order
derivatives on all the fields. For $P_\mu$ and $M_{\mu\nu}$ there are
the usual requirements of a covariant action. For the local
dilatations we have the requirement that the weights of all fields in
each term should add up to $d$, where $\partial_\mu$ counts also for
1, as can be seen from \eqn{delCderphi}. Indeed, the explicit $\Lambda_D$
transformations finally have to cancel with
\begin{equation}
\xi^\mu(x)\partial_\mu{\cal L}\approx -(\partial_\mu \xi^\mu (x)){\cal
L}=-d\Lambda_D(x){\cal L}\,.
\end{equation}
For special conformal transformations one remains with
\begin{equation}
\delta_K S=2\Lambda_K^\mu  \int d^dx\,
\frac{{\cal L}\dr}{\partial(\partial_\nu\phi^i)}
\left(- \eta_{\mu\nu}w_i\phi^i +2m_{\mu\nu}{}^i{}_j\phi^j \right)
+\Lambda_K^\mu\frac{ S\delr}{\delta \phi^i(x)}(k_\mu\phi)^i(x)\,.
\label{deltaLK}
\end{equation}
where $\dr$ indicates a right derivative. The first terms originate
from the $K$-transformations contained in \eqn{ximu} and \eqn{Lambdax}.
In most cases these are sufficient to find the invariance and no
$(k_\mu\phi)$ are necessary. In fact, the latter are often excluded
because of the requirement that they should have Weyl weight $w_i-1$,
and in many cases there are no such fields available.
\par
Although we will show that this condition is satisfied for many
dilatational invariant theories, it is non-trivial. As a
counterexample we give the action of the scalars $\phi^1$ and
$\phi^2$ (with Weyl weights $(\ft d2-1)$)
\begin{equation}
{\cal L}=\left( 1+\frac{\phi^1}{\phi^2}\right)
(\partial_\mu\phi^1)  (\partial^\mu\phi^2)\,.
\end{equation}
\noindent \textit{Exercise.}

There are typical cases in which \eqn{deltaLK} does not receive any
contributions. Check the following ones
\begin{enumerate}
\item scalars with Weyl weight 0.
\item spinors appearing as $\dsl\lambda$ if their Weyl weight is
$(d-1)/2$. This is also the appropriate weight for actions as
$\bar \lambda  \dsl\lambda $.
\item \label{exampleconf}
Vectors or antisymmetric tensors whose derivatives appear only as
field strengths
$ \partial_{[\mu_1}B_{\mu_2\ldots \mu_p]}$ if their Weyl weight is
$p-1$. This value of the Weyl weight is what we need also in order
that their gauge invariances and their zero modes commute with the
dilatations. Then scale invariance of the usual square of the field
strenghts will fix $p=\ft d2$.
\item Scalars $X^i$ with Weyl weight $\ft d2-1$ and
\begin{equation}
{\cal L}=  (\partial_\mu X^i) A_{ij} (\partial^\mu X^j)  \,,
\label{scalarrigidaction}
\end{equation}
where $A_{ij}$ are constants.
\end{enumerate}

\subsection{Gauge theory and gravity}\label{ss:gravity}

{}From the theory of general relativity we know how to construct actions
invariant under the local Poincar\'{e} group. However, I will now
consider the constructions from an algebraic viewpoint, i.e.\ as a
gauge theory of the \Poin\ group. First, recall the general equations for
gauge theories with infinitesimal transformations
\begin{equation}
\delta (\epsilon )=\delta _A(\epsilon ^A)=\epsilon^A T_A\,,
\end{equation}
where $A$ thus labels all the symmetries, and $\epsilon^A$ are all the
parameters. For every symmetry one introduces a gauge field $h_\mu^A$, and
if the algebra is
\begin{equation}
[\delta_A( \epsilon_1^A),\delta_B(\epsilon_2^B)] =\delta_C\left(
\epsilon_2^B\epsilon_1^A f_{AB}{}^C\right) \,,
\end{equation}
then these transform as
\begin{equation}
\delta(\epsilon) h_\mu^A=
\partial_\mu\epsilon^A +\epsilon^C h_\mu^B f_{BC}{}^A\,.
\label{delGaugef}
\end{equation}
Covariant derivatives are defined as
\begin{equation}
\nabla_\mu =\partial_\mu -\delta_A(h^A_\mu)\,,
\end{equation}
and their commutators are new transformations with as parameters the
curvatures:
\begin{eqnarray}
[\nabla_\mu,\nabla_\nu]&=&-\delta_A(R^A_{\mu\nu})\nonumber\\
 R_{\mu\nu}^A &=& 2\partial_{[\mu}h_{\nu]}^A+ h_\nu^C h_\mu^B f_{BC}{}^A\,,
 \label{defcurv}
\end{eqnarray}
which transform `covariantly' as
\begin{equation}
\delta  R_{\mu\nu}^A = \epsilon^C R_{\mu\nu}^B f_{BC}{}^A\,. \label{deltacurv}
\end{equation}
\par
I now apply this to the \Poin\ group with gauge fields
\begin{equation}
h_\mu^A T_A= e_\mu^a P_a +\omega_\mu{}^{ab}M_{ab}\,.
\end{equation}
The indices $\mu, \nu, \ldots $ of
the algebra of the previous sections should now be replaced by flat indices
$a, b, \ldots $. The curvatures are then (for later convenience I put an
indication $P$ for the \Poin\ curvatures)
\begin{eqnarray}
R^{P}_{\mu\nu}(P^a)&=&2\partial_{[\mu}e_{\nu]}^a+2\omega_{[\mu}{}^{ab}e_{\nu] b}
\nonumber\\
R^P_ {\mu\nu}(M^{ab})&=&2\partial_{[\mu}\omega_{\nu]}{}^{ab}
+2\omega_\mu^{c[a}\omega_\nu{}^{b]}{}_c \,.
\end{eqnarray}
Usually the spin connection is not considered as an independent
field. One can obtain the relation between the spin connection and
the vierbeins by imposing a constraint
\begin{equation}
R^P_{\mu\nu}(P^a) =0 \,.  \label{constrRP}
\end{equation}
As the `vielbein' $e_\mu^a$ is assumed to be invertible, this
 can be solved for the `spin connection'
\begin{equation}
\omega_\mu{}^{ab}=2e^{\nu[a}\partial_{[\mu}e_{\nu]}{}^{b]}
-e^{a\rho}e^{b\sigma}
e_{\mu c}\partial_{[\rho}e_{\sigma]}^c \,.  \label{omegamuab}
\end{equation}
Such a constraint, which can be solved for a field, is called
{\it a conventional constraint}. Similar constraints are often used in the
superspace approach.

The constraint is not invariant under all the symmetries. Therefore the theory
with the constraint, and thus a dependent spin connection has an
algebra that is different from (\ref{Poinalgebra}). Essentially
the translations $P_a$ are replaced by `covariant general coordinate
transformations'
\begin{equation}
\delta_{cgct}(\xi)=\delta_{gct}(\xi)-\delta_I(\xi^\mu h_\mu^I)\,,
\end{equation}
where $I$ stands for all transformations except the translations. In
fact, one can check that the translations $P_a$ on the vierbein take the
form of covariant general coordinate transformations, due to the
constraint \eqn{constrRP}.

{}From now on we replace  translations
by these covariant general coordinate transformations. On non-gauge fields
we have (and we write these equations again more general than for the pure
\Poin\ algebra, as we want to use it later for larger algebras containing
translations)
\begin{equation}
\nabla_\mu \phi=\partial_\mu\phi -e_\mu^a P_a \phi - \delta_I(h_\mu^I)\phi=0\,,
\label{nablaphi=0}
\end{equation}
which, solved for $P_a$, gives what is now called the covariant derivative
\begin{equation}
P_a\phi= D_a\phi\equiv e_a^\mu\left(\partial_\mu\phi - \delta_I(h_\mu^I)\phi
\right) \,. \label{defcovder}
\end{equation}
While in the original algebra translations did commute, (covariant)
general coordinate transformations do not commute, i.e.\ $f_{ab}^A\neq
0$. If one takes this into account in \eqn{defcurv}, thus adding a new term
with $f_{ab}^A e^a_\mu e^b_\nu$, then the full curvature
vanishes, consistent with \eqn{nablaphi=0}. Then one can solve this again for
$f_{ab}^A$, and one obtains
\begin{equation}
 f_{ab}^A =-e_a^\mu e_b^\nu R_{\mu\nu}^A \equiv -R_{ab}^A\,,
\end{equation}
such that still
\begin{equation}
[D_a,D_b]=-\delta_I( R^I_{ab})\,.
\end{equation}
In the curvature formula \eqn{defcurv} one thus takes $f_{ab}^I=0$, as
 in the original algebra, but the other sums over symmetries include also the
translations.

The Ricci tensor is a contraction of the $M$-curvature
\begin{equation}
R_{\mu\nu}=  R^P_{\mu\rho}(M^{ba})e_b{}^\rho e_{\nu a}\,, \qquad
R=R_{\mu\nu}g^{\mu\nu}\,, \label{Rconv}\end{equation}
and the field equation of the usual pure \Poin\ action is
\begin{equation}
\frac{\delta }{\delta g^{\mu\nu} }\int d^4x\sqrt{-g}\,R =
\sqrt{-g}\,\left(R_{\mu\nu}-\ft12 g_{\mu\nu}R\right)\,.
\end{equation}

In the rigid theory, a scalar does transform under Lorentz rotations. Indeed, see
\eqn{deltaC} where $\xi(x)$ contains a Lorentz rotation according to \eqn{ximu}.
This is not the case in the local theory. Here,
$\xi(x)$ is the local parameter for general coordinate transformations.
The former rotation is thus part of the
general coordinate transformation. In the local theory,
scalars are invariant under Lorentz rotations. For other fields, the rotations
exclusively
come from the part in $m_{ab}$ (and in a conformal theory, the full $\Lambda_M(x)$
is the local parameter, and the special conformal transformations are not
part of it, but are exclusively contained in $(k_\mu\phi)$). E.g.\
for a spinor, $\psi$, the covariant derivative is
(if they do not transform under any other symmetry)
\begin{equation}
D_a\psi= e_a^\mu
\left(\partial_\mu+\ft14\omega_\mu{}^{bc}\gamma_{bc}\right)\psi \,.
\label{Dapsi}
\end{equation}
Further it is important to realize that the rotations $M^{ab}$ act on
vectors $V_a$, but a local spacetime vector $V_\mu$ is invariant. On
the other hand, the general coordinate transformation on $V_\mu$ has
not just the $\xi ^\nu \partial _\nu V_\mu $ term, but also the term
$V_\nu \partial _\mu \xi ^\nu$.

The following theorem is extremely useful when calculating transformations
\cite{Karpacz2,d6SC}.

\noindent \textbf{Theorem on covariant derivatives.}
\textit{We define a covariant quantity as one whose transformation has no
derivative on a parameter. In particular, it has to be a world scalar
such that general coordinate transformations on the field do not
contain a derivative on $\xi ^\mu $. If a
covariant quantity transforms only into covariant quantities, then
a covariant derivative \eqn{defcovder} on such a covariant quantity is a new
covariant quantity.} 
\par
\noindent
This implies that transformations of a covariant
quantity contain only derivatives as covariant derivatives $D_a$ on
other covariant quantities, or in the form of curvatures with Lorentz
indices $R_{ab}$.

\noindent \textbf{Proof}. In this proof we will allow the
algebra to be `open', see below. First, we give the conditions again
 in formulas, which may
clarify some of the sentences in the statement of the proof.
Recall that the full set of generators (which correspond to
indices $A,B,\ldots $), contain translations (indices $a,\ldots $) and
the remaining transformations are indicated by indices $I$.
Gauge fields of these other symmetries may have transformation laws
with other terms than in (\ref{delGaugef}), in particular with matter
fields. Their transformations under the (non-translation)
symmetries are
\begin{equation}
  \delta_I(\epsilon^I ) h_\mu ^I=\partial _\mu \epsilon ^I+\epsilon ^J h_\mu ^A
  f_{AJ}{}^I +\delta _{m}h_\mu ^I\,.
\label{delGaugefm}
\end{equation}
The last term can contain `covariant' matter fields, but not explicit
gauge field (apart from $e_\mu ^a$). Consider a matter field $\phi $ that
transforms as
\begin{equation}
  \delta_I(\epsilon^I ) \phi =\epsilon ^I (T_I \phi) \,,
\label{deltaphi}
\end{equation}
which has thus no derivative on $\epsilon ^I$. The requirement that a
covariant quantity transforms in a covariant quantity is the
statement that
\begin{equation}
  \delta_J(\epsilon^J ) (T_I\phi) =\epsilon ^J(T_JT_I\phi)\,.
\label{delTIphi}
\end{equation}
First, consider the transformation of the covariant derivative
\begin{equation}
  D_\mu \phi =\partial _\mu \phi -h_\mu ^I T_I\phi \,.
\label{Dmuphi}
\end{equation}
The algebra is
\begin{equation}
  \left[ \delta (\epsilon _1),\delta (\epsilon _2)\right] \phi =
  \epsilon _2^J\epsilon ^I_1\left[ T_I,T_J\right\} \phi =
\epsilon _2^J\epsilon ^I_1\left( f_{IJ}{}^{K}T_K\phi
+f_{IJ}{}^a D_a\phi  +\eta
_{IJ}\right)\,.
\label{algebraphi}
\end{equation}
Remark that all formulas have been written in a way that they are
also applicable to fermionic transformations (thus the order in which
fields and parameters are written, is carefully chosen). The symbol
$[.,.\}$ will be introduced later in (\ref{defcac})
The commutator can include
general coordinate transformations, which act on the $\phi $ as $D_a\phi
=e_a^\mu D_\mu \phi $. Furthermore,
 $\eta $ is a possible non-closure term.
The previous formulas imply
\begin{eqnarray}
\delta(\epsilon )  D_\mu \phi & = & \epsilon ^I\partial _\mu T_I\phi -
\epsilon ^J h_\mu ^A
  f_{AJ}{}^I T_I\phi-(\delta _{m}h_\mu ^I)T_I\phi
  -h_\mu ^I\epsilon ^J(T_JT_I\phi) \nonumber\\
& = & \epsilon ^I\left(\partial _\mu-h_\mu ^JT_J\right)T_I\phi
-\epsilon ^Je_\mu ^af_{aJ}{}^IT_I\phi \nonumber\\
&&-h_\mu ^I\epsilon ^Jf_{JI}{}^aD_a\phi -h_\mu ^I\epsilon ^J\eta _{JI}
-(\delta _{m}h_\mu ^I)T_I\phi\,.
\label{delDmuphi1}
\end{eqnarray}
Consider now the transformation of (\ref{defcovder}), using
\begin{equation}
  \delta e_a^\mu=-e_b^\mu e_a^\nu \delta e_\nu ^b=-\epsilon ^Ih_a^A
  f_{AI}{}^b e_b^\mu\,.
\label{deleinv}
\end{equation}
Then the first term on the last line of (\ref{delDmuphi1}) is
cancelled and we obtain
\begin{equation}
  \delta(\epsilon ) D_a \phi= \epsilon ^ID_aT_I\phi -\epsilon ^If_{aI}{}^A T_A \phi
  +\epsilon ^Ih_a^J\eta _{JI}-(\delta _{m}h_a ^I)T_I\phi\,.
\label{delDaphi}
\end{equation}
\QED
Consider the final result. In the second term the index $A$ is used, which
may take the value $a$, with $T_a=D_a$. Otherwise, this term appears whenever a
gauge field $h_\mu ^I$ has in its transformation law a term which is
proportional to the vielbein $e_\mu ^a$. As in our terminology on
covariant quantities we did not consider the vierbein as a gauge
field, this term can be included in $\delta _m h_\mu ^I$. Thus for
closed algebras ($\eta _{JI}=0$) the result is as follows: in the
transformation of the covariant derivative, the derivative should not
work on the parameter, and further terms appear only from the
transformation of gauge fields that are not proportional to gauge
fields.
Thus gauge fields never appear `naked' in such transformations. One only has to
be careful here that the vielbein is not considered as a gauge field itself.

E.g.\ consider the calculation of the transformation of \eqn{Dapsi}. First of
all, remark that this would not work for $D_\mu\psi$, which is not a
covariant quantity. Then, the transformations of the spin connection
$\omega_\mu^{bc}$ can be neglected as this field transforms only in the
derivative of a parameter or in an explicit gauge field, and the latter do not
appear in the final result due to the theorem. We do have to take into account
that the vielbein transforms as
\begin{equation}
\delta e_\mu^a= \xi^\nu\partial_\nu e_\mu^a -\lambda_M^{ab}e_{\mu b}+\ldots \,,
\end{equation}
where the $\ldots $ stand for e.g.\ a term $e_\nu^a\partial_\mu \xi^\nu$, which
can be neglected because it contains a derivative on the parameter.
The final result is thus
\begin{eqnarray}
\delta D_a\psi&=& \left( \xi^\nu\partial_\nu e^\mu_a
-\lambda_{M\,ab}e^\mu_b\right) \partial_\mu\psi + e_\alpha^\mu\partial_\mu
\left( \xi^\nu\partial_\nu  -\ft14 \lambda_M^{bc}\gamma_{bc}\right)
\psi+\ldots \nonumber\\
&=& \xi^b D_b D_a\psi - \lambda_M^{ab} D_b\psi -\ft14 \lambda_M^{bc}\gamma_{bc}
D_a\psi\,.
\end{eqnarray}
This shows the usefulness of the theorem:
although in intermediate steps we omitted some terms, at the end we
obtain the full result.

\subsection{Local conformal transformations}\label{ss:localconf}

We now consider the gauging of the conformal algebra, and show how Poincar\'{e}
gravity is recovered in a gauge for dilatations and special conformal
transformations. The gauge fields are introduced as
\begin{equation}
h_\mu^A T_A= e_\mu^a P_a +\omega_\mu{}^{ab}M_{ab}+b_\mu D +f_\mu^a K_a\,.
\end{equation}
The curvatures are
\begin{eqnarray}
R_{\mu\nu}(P^a)&=& R^P_{\mu\nu}(P^a)+2b_{[\mu}e_{\nu]}^a  \nonumber\\
R_{\mu\nu}(M^{ab})&=&R^P_{\mu\nu}(M^{ab})+8f_{[\mu}^{[a}e_{\nu]}^{b]}
\nonumber\\
R_{\mu\nu}(D)&=&2\partial_{[\mu}b_{\nu]}-4f_{[\mu}^{a}e_{\nu]a}\nonumber\\
R_{\mu\nu}(K^a)&=&2\partial_{[\mu} f_{\nu]}^a +2\omega_{[\mu}{}^{ab}f_{\nu] b}
-2b_{[\mu}f_{\nu]}^a \,.
\end{eqnarray}
The conventional constraint \eqn{constrRP} could be imposed because the gauge
field $\omega_\mu^{ab}$ appears in the curvature multiplied by the invertible
vielbein. Considering the curvatures, one can see that also the gauge field
$f_\mu^a$ appears in such a way in the curvatures of $D$ and $M$. It can be
solved if we impose as second conventional constraint:
\begin{eqnarray}
&&  R_{\mu\nu}(P^a) =0 \nonumber\\
&& R_{\mu\nu}(M^{ab})e_b^\nu=0\,.
\end{eqnarray}
Therefore as well $\omega_\mu^{ab}$ as $f_\mu^a$ are dependent
fields. In the first one there is a  modification to \eqn{omegamuab}
due to the term proportional to $b_\mu$. The solution for the
$K$ gauge field is  (with the definitions as in (\ref{Rconv}))
\begin{equation}
f_\mu^a =\frac1{2(d-2)} R_\mu{}^a-\frac1{4(d-1)(d-2)}R e_\mu^a\,.  \label{solfmua}
\end{equation}

Because of the theorem on covariant derivatives it is useful
to give the part of the transformations of gauge fields proportional
to vielbeins. These are (with the dots representing the
transformations in other explicit gauge fields)
\begin{eqnarray}
\delta e_\mu^a&=&-\Lambda_D e_\mu^a -\Lambda_M^{ab}e_{\mu b}+\ldots \nonumber\\
\delta b_\mu &=& 2\Lambda_K^a e_{\mu a}+\ldots\nonumber\\
\delta \omega_\mu^{ab}&=&- 4\Lambda_K^{[a}e_\mu^{b]}+\ldots\,.\label{trgaugeine}
\end{eqnarray}

\noindent \textit{Exercise}

It is interesting to compare transformations of derivatives of
fields in the rigid theory with those of covariant derivatives in
conformal gravity. In \eqn{delCderphi} the second line came from the
derivatives of $\xi$ (first two terms) and the derivative of
$\Lambda_D$. In the gauge theory, the full $x$-dependent $\xi _\mu(x) $
is the parameter of general coordinate transformations. Show, with the
help of the theorem on covariant derivatives, how one re-obtains
(\ref{delCderphi}) with all derivatives replaced by covariant ones
(and changing to $a$-indices).

Show that also in the local theory, (\ref{deltaLK}) is the rule
to determine whether the action with ordinary derivatives replaced by
covariant derivatives, is an invariant. \QED
The transformation law of a covariant derivative determines the covariant box
\begin{eqnarray}
\bbox^C \phi& \equiv &\eta^{ab}D_b D_a \phi=e^{a\mu}\left( \partial_\mu
D_a \phi -(w+1) b_\mu D_a \phi +\omega_{\mu\,ab}D^b\phi+2wf_{\mu a}\phi \right) \nonumber\\
&=& e^{-1}\left( \partial_\mu-(w+2-d)b_\mu\right)e g^{\mu\nu}\left(
\partial_\nu-wb_\nu\right) \phi+\frac{w}{2(d-1)}R\,\phi \,.
\end{eqnarray}
The latter is the well-known $R/6$ term in $d=4$. In fact, choosing
$w=\frac{d}{2}-1$, one has a conformal invariant scalar action
\begin{equation}
I=\int d^d x\, e \phi\bbox^C \phi \,. \label{Iscalarconf}
\end{equation}
\noindent \textit{Exercise}: show that $\int d^d x\, e D_a\phi D^a
\phi$ is not invariant. How does the difference between this
statement and the one in (\ref{scalarrigidaction}) occur?

Consider now that we want a \Poin\ invariant action. Then we have to
break dilatations and special conformal transformations, as these are
not part of the \Poin\ algebra.
Considering \eqn{trgaugeine} it is clear that the latter can be
broken by a gauge choice
\begin{equation}
K-\mbox{gauge :}\qquad b_\mu=0\,.
\end{equation}
Therefore, of the `Weyl multiplet' (the multiplet of fields with the
gauge fields of the conformal algebra), only the vielbein remains.
One could take as gauge choice for dilatations a fixed value of
a scalar $\phi$. And one easily checks that then the action \eqn{Iscalarconf}
reduces to the \Poin\ gravity action.

\section{Spinors in arbitrary dimensions}\label{ss:spinors}
I now present information about the Clifford algebra and spinors in
arbitrary dimensions. I will discuss arbitrary signatures of
spacetime. Different from many other treatments, I will not consider
arguments of field theory.
E.g.\  in many other papers one uses that spinors are solutions of the
Dirac equation. That is also the case in \cite{Scherk}, which is however the first
reference for the facts that I will recall.
\subsection{Gamma matrices and their symmetry}
I consider arbitrary spacetime dimensions $d=t+s$, with $t$ timelike
directions and $s$ spacelike directions, and consider the general facts of
Clifford algebras and spinors, as has been done first in \cite{KugoTown}.
\par
The Clifford algebra is
\begin{equation}
\Gamma_a \Gamma_b + \Gamma_b \Gamma_a = 2\eta_{ab}\,,
\label{defClifford}
\end{equation}
where $\eta=\diag (-\ldots -+\ldots +)$, writing first the timelike
directions and then the spacelike ones.

I first give a representation of the Clifford algebra for signature
$(0,d)$ (only spacelike) in terms of $\sigma$ matrices
\begin{equation}
 \sigma_1=\left(\begin{array}{cc} 0&1\\1&0\end{array}\right )\hspace{0.5cm}
\sigma_2=\left(\begin{array}{cc} 0&-\rmi\\ \rmi&0\end{array}\right
)\hspace{0.5cm} \sigma_3=\left(\begin{array}{cc}
1&0\\0&-1\end{array}\right )\,.
\end{equation}
The only relevant properties are
$\sigma_1\sigma_2=\rmi\sigma_3$ and cyclic, and that they square to
$\unity _2$ and are hermitian.
\begin{eqnarray}
\Gamma_1&=&\sigma_1\otimes \unity \otimes \unity \otimes \ldots
\nonumber\\ \Gamma_2&=&\sigma_2\otimes \unity \otimes \unity \otimes
\ldots \nonumber\\ \Gamma_3&=&\sigma_3\otimes \sigma_1 \otimes \unity
\otimes \ldots \nonumber\\ \Gamma_4&=&\sigma_3\otimes \sigma_2
\otimes \unity \otimes \ldots \nonumber\\ \Gamma_5&=&\sigma_3\otimes
\sigma_3 \otimes \sigma_1 \otimes \ldots \nonumber\\ \ldots  &=&
\ldots \,.
\label{representationClifford}
\end{eqnarray}
This is for even dimensions a representation of dimension $2^{d/2}$. For odd
dimensions, e.g.\ $d=5$, one does not need the last $\sigma_1$ factor in
$\Gamma_5$, and the dimension is still $2^{(d-1)/2}$. These matrices are all
hermitian. If there are $t$ timelike directions, we just have to multiply the
first $t$ matrices in the representation above by $\rmi$. E.g.\ for
the Minkowski case $\Gamma _1=\rmi\sigma _1\otimes \unity \otimes \ldots $.
 Thus we get as hermiticity property for timelike
and spacelike directions
\begin{equation}
\Gamma_t^\dagger =-\Gamma_t\,,\qquad \Gamma_s^\dagger =\Gamma_s\,,
\label{hermitianGamma}
\end{equation}
or in general
\begin{equation}
\Gamma_a^\dagger =(-)^t A\Gamma_a A^{-1}\,,\qquad A=\Gamma_1\ldots
\Gamma_t\,.\label{Gammadagger}
\end{equation}
Of course the realization given above is not unique. One preserves
(\ref{defClifford}) with
\begin{equation}
\Gamma'=U^{-1}\Gamma U\,.
\label{GammaUnittransf}
\end{equation}
To respect also (\ref{hermitianGamma}), $U$ has to be unitary. A
further remark on the uniqueness will follow in
footnote~\ref{fn:uniqCl}.
\par
Another important fact is that {\bf in even dimensions} a complete
set of $2^{d/2} \times 2^{d/2}$ matrices is provided by $\{
\Gamma^{(n)}\}$ with $n=0,1,\ldots d$, and
\begin{equation}
\Gamma^{(n)}=\Gamma_{a_1\ldots a_n}\equiv
\Gamma_{[a_1}\Gamma_{a_2}\ldots \Gamma_{a_n]}\,.
\end{equation}
The last matrix $\Gamma^{(d)}$, we write as\footnote{If $0$ is used
as label of the time direction in Minkowski space, then we have $\Gamma_*=
(-\rmi)^{d/2+1}\Gamma_0\Gamma_1\ldots \Gamma_{d-1}$.}
\begin{equation}
\Gamma_*=(-\rmi)^{d/2+t}\Gamma_1\ldots \Gamma_d\,, \qquad
\Gamma_* \Gamma_* =1\,,
\label{Gamma*}
\end{equation}
where the normalization is chosen such that the last equation holds
in any case. $\Gamma _*$  is independent on the signature
of spacetime, in the sense that when I introduced first the $\Gamma $
matrices in Euclidean signature, and then went to an arbitrary
signature, I multiplied the first $t$ gamma matrices by $\rmi$. This
drops out again by the multiplication by $(-\rmi)^t$ in (\ref{Gamma*}).
In the representation (\ref{representationClifford}), $\Gamma_*$ is
\begin{equation}
  \Gamma_*=\sigma_3\otimes  \sigma_3\otimes
\sigma_3\otimes  \ldots\,.
\label{reprGammaStar}
\end{equation}
For even dimensions we have (the Levi--Civita tensor $\varepsilon $
is defined in appendix~\ref{app:conv})
\begin{equation}
\Gamma_{a_1\ldots a_n}=\frac{1}{(d-n)!}
\varepsilon_{a_1\ldots a_d}\rmi^{d/2+t}\Gamma_*\Gamma^{a_d\ldots a_{n+1}}\,.
\label{Gamma*Gamma}
\end{equation}
Note that $\Gamma_*$ anticommutes with $\Gamma_a$ and can thus
 be used as $\Gamma_{d+1}$
in the next, {\bf  odd dimension}. In odd dimensions the product of
all $\Gamma$ matrices gives $\pm 1$ or $\pm \rmi$, and a basis is formed
by $\Gamma^{(n)}$ with $n=0,1,\ldots , (d-1)/2$.
\par
 There is always a `{\bf charge conjugation
matrix}' ${\cal C}$, such that
\begin{equation}
{\cal C}^T=-\varepsilon {\cal C}\,, \qquad \Gamma_a^T=-\eta{\cal C}\Gamma_a
{\cal C}^{-1}\,,   \label{calC}
\end{equation}
for $\varepsilon=\pm 1$ and $\eta=\pm 1$. A formal proof can be found
in \cite{Scherk,KugoTown}, but I prove it here by giving two
possibilities in the representation (\ref{representationClifford})
for even dimensions
\begin{equation}
\begin{array}{rcll}
 {\cal C}_+&=& \sigma_2\otimes
\sigma_1\otimes \sigma_2 \otimes \sigma_1\otimes \ldots\qquad&
\eta =+1
\nonumber\\
{\cal C}_-&=&  \sigma_1\otimes \sigma_2 \otimes \sigma_1\otimes
\sigma_2\otimes \ldots \propto {\cal C}_+ \Gamma_*\qquad&\eta =-1 \,.
\end{array}
\label{Crepresentation}
\end{equation}
Note that they satisfy ${\cal C}={\cal C}^\dagger ={\cal C}^{-1}$.
For odd dimensions, only one of the two can be
used, see  table~\ref{tbl:epseta} below.
If one changes the representation as in (\ref{GammaUnittransf}),
then, to preserve (\ref{calC}),
the charge conjugation matrix should transforms as
\begin{equation}
  {\cal C}'=U^T {\cal C}U\,.
\label{CUnittransf}
\end{equation}
Therefore you see that the unitarity, ${\cal C}^\dagger ={\cal C}^{-1}$,
 is true in any representation, but  the charge conjugation not
necessarily squares to $\unity $.
\par
We will now deduce which signs of the two variables $\epsilon $ and
$\eta $ are allowed for
each value of $s$ and $t$. First remark that
\begin{equation}
\left({\cal C}\Gamma^{(n)}\right) ^T= -\epsilon(-)^{n(n-1)/2}(-\eta)^n
{\cal C}\Gamma^{(n)}\,. \label{symGamman}
\end{equation}
This has a periodicity of $n\rightarrow n+4$. This periodicity
implies that the number of symmetric matrices should be the sum of
lines of the following formulae
\begin{eqnarray}
{d\choose 0} +{d\choose 4} +\ldots &=&2^{d-2}+2^{d/2-1}\cos\frac{d\pi}{4}\nonumber\\
{d\choose 1} +{d\choose 5} +\ldots &=&2^{d-2}+2^{d/2-1}\sin\frac{d\pi}{4}\nonumber\\
{d\choose 2} +{d\choose 6} +\ldots &=&2^{d-2}-2^{d/2-1}\cos\frac{d\pi}{4}\nonumber\\
{d\choose 3} +{d\choose 7} +\ldots &=&2^{d-2}-2^{d/2-1}\sin\frac{d\pi}{4}\,.
\label{d0123}
\end{eqnarray}
 On the other hand we can count
that  there should be
\begin{eqnarray}
2^{\left[{\rm Int ~}\frac{d}{2}\right] -1}
\left( 2^{\left[{\rm Int ~}\frac{d}{2}\right]}+1\right) &   & \mbox{symmetric matrices} \nonumber\\
2^{\left[{\rm Int ~} \frac{d}{2}\right]-1}
\left( 2^{\left[{\rm Int ~} \frac{d}{2}\right]} -1\right) &   &  \mbox{antisymmetric matrices.}
\label{ntsymmasymm}
\end{eqnarray}
For odd dimensions, there is a unique solution. For even dimensions, two of
the lines in (\ref{d0123}) are equal, and there are thus two
possibilities for which matrices are symmetric. This corresponds to
the two possibilities for the charge conjugation matrix in that case.
The result is given in table~\ref{tbl:epseta}.
\begin{table}[ht]
\begin{center}\begin{tabular}{||c|cc|cc|}\hline
d (\mbox{mod }8)& S & A  &$\epsilon$&$\eta$\\ \hline\hline
0           & 0,3& 2,1   &$-1$   &$+1$\\
            & 0,1 & 2,3  &$-1$   &$-1$\\ \hline
1           & 0,1 & 2,3  &$-1$   &$-1$\\      \hline
2           & 1,0  & 3,2 &$-1$   &$-1$\\
            & 1,2  & 3,0 &$+1$   &$+1$\\  \hline
3           & 1,2 & 0,3  &$+1$   &$+1$\\       \hline
4           & 2,1  & 0,3 &$+1$   &$+1$\\
            & 2,3  & 0,1 &$+1$   &$-1$\\  \hline
5           & 2,3 & 0,1  &$+1$   &$-1$\\      \hline
6           & 3,2  & 1,0 &$+1$   &$-1$\\
            & 3,0  & 1,2 &$-1$   &$+1$\\  \hline
7           & 0,3 & 1,2  &$-1$   &$+1$\\ \hline\hline
\end{tabular}\caption{\sl For all dimensions (modulo 8) one finds which
${\cal C}\Gamma^{(n)}$ should be symmetric (S) and antisymmetric (A).
These values of $n$ are modulo 4). This determines signs of
$\epsilon$ and $\eta$.
}\label{tbl:epseta}\end{center}\end{table}

\noindent \textit{Exercise}

Check with the explicit representation the signs of $\eta $ and
 $\epsilon $ for the two possible charge conjugation matrices in 4
 dimensions. Then consider the fifth gamma matrix, and see that the
 (\ref{calC}) only holds for one choice of $ {\cal C}$.
\subsection{Irreducible spinors}
We still do not know whether the spinor is irreducible. There are two
types of projections that one can envisage. The first is only for even
dimensions, where we have $\Gamma_*$ available as in \eqn{Gamma*}. We can
thus define left and right chiral (or `Weyl') spinors. For definiteness
we take
\begin{equation}
\lambda_L=\ft12\left(1+\Gamma_*\right) \lambda \,,\qquad
\lambda_R=\ft12\left(1-\Gamma_*\right) \lambda\,.
\label{LRspinors}
\end{equation}
E.g.\ in four dimensional Minkowski space ($s=3,t=1$) we define%
\footnote{For 4 dimensional Minkowski space we write $\gamma _0$,\ldots ,
$\gamma _3$ in stead of $\Gamma _1$,\ldots ,$\Gamma _4$.}
\begin{equation}
\gamma_5=\Gamma _* =\rmi\gamma_0\gamma_1\gamma_2\gamma_3\,, \qquad
\lambda_L =\ft12 (1+\gamma_5) \lambda\,, \qquad
\lambda_R =\ft12 (1-\gamma_5) \lambda\,.
\label{defgamma5}
\end{equation}

The second possible projection is a \textit{reality condition}.
Consider first the reality properties of the $\Gamma$ matrices.
Combining \eqn{Gammadagger} and \eqn{calC}, we have\footnote{Here we
can give some remark on the uniqueness of the Clifford algebra representation.
 It is clear that $\pm \Gamma ^*$ can also be used as representations of
 the Clifford algebra. The equation (\ref{Gammacc}), which for even dimensions
 holds for two possibilities of $\eta $, implies that in even dimensions these
 representations are unitary equivalent with the original one. The fact that for
 odd dimensions only one sign of $\eta $ can be used implies that in odd dimensions
 there are two unitary inequivalent representations.\label{fn:uniqCl} }
\begin{equation}
\Gamma_a^*=-\eta(-)^t B\Gamma_a B^{-1}\,,\qquad B^T={\cal C}A^{-1}\,.
\label{Gammacc}
\end{equation}
As  $A$ and ${\cal C}$ are unitary, we  find that also
$B$ is unitary. One obtains
\begin{equation}
B^* B= -\epsilon\eta^t(-)^{t(t+1)/2}\,. \label{B*B}
\end{equation}
To prove this equation, combining (\ref{Gammadagger}) and~(\ref{calC})
implies that
\begin{equation}
  A^T=\Gamma _t^T \cdots \Gamma _1^T= (-\eta )^t{\cal C}\Gamma _t\cdots
\Gamma _1=\eta ^t{\cal C}A^{-1}{\cal C}^{-1}=
  (-)^{t(t+1)/2}\eta ^t {\cal C}A{\cal C}^{-1}\,.
\label{AT}
\end{equation}
\par
Let us now try to put a reality condition on spinors
\begin{equation}
\lambda^*= \tilde B \lambda\,, \label{realcond}
\end{equation}
for some matrix $\tilde B$. First of all we want consistency with Lorentz
transformations, i.e.\ taking the Lorentz transformation of both sides
\begin{eqnarray}
\left( -\ft14 \Gamma_{ab} \lambda\right)^*&=& -\ft14 \tilde B \Gamma_{ab}\lambda
\qquad \Rightarrow
\nonumber\\
B \Gamma_{ab} B^{-1}\tilde B &=& \tilde B \Gamma_{ab}  \,,
\end{eqnarray}
and therefore we take
\begin{equation}
  \tilde B=\alpha B\,.
\label{Bprimealpha}
\end{equation}
For consistency, $\lambda^{**}=\lambda$, \eqn{realcond} implies that
$\tilde B^*\tilde B=1$. Combining this with \eqn{B*B} gives $|\alpha|=1$
and \textbf{the right hand side of \eqn{B*B} should be~1}. It turns out
that this condition is satisfied for\footnote{I thank I. Masina for
pointing out a sign mistake in the previous version of this review.}
\begin{eqnarray}
s-t&=&0,1,7\ \mod 8 \nonumber\\
s-t&=&2\ \mod 8 \ \mbox{ with  }\eta(-)^{d/2}=+1\nonumber\\
s-t&=&6\ \mod 8 \ \mbox{ with  }\eta(-)^{d/2}=-1\,. \label{dstMajorana}
\end{eqnarray}
In these cases, we can thus define a reality condition as in \eqn{realcond},
and such spinors will be denoted as `Majorana spinors'.
Another way of expressing the Majorana condition is that the
`Majorana conjugate' equals the `Dirac conjugate'. They are
respectively
\begin{equation}
\bar \lambda \equiv \lambda ^T{\cal C}\,,\qquad
\bar \lambda^C\equiv \lambda^\dagger A\alpha^{-1}\,.    \label{Majorana}
\end{equation}
Majorana spinors can thus be thought as spinors $\lambda_1 + \rmi \lambda_2$,
where $\lambda_1$ and $\lambda_2$ have real components, but these are
related by the above condition.
The exact value $\alpha$, if it is of modulus 1, is irrelevant for
our purpose here. However, it determines whether e.g.\ $\bar \psi \psi $
or $\int d^d x \bar \psi\not\!\!\partial \psi $ is hermitian. We will
come back to this issue in section~\ref{ss:remarksCl}.

Consider now that the right had side of \eqn{B*B} is not 1, thus $-1$.
Then there is still another possibility if we have extended
supersymmetry. Indeed, then one can define a `\textit{symplectic Majorana
condition}'
\begin{equation}
\lambda^*_i\equiv \left( \lambda ^i\right) ^*= B \Omega_{ij}\lambda^j\,,
\label{symplMaj}
\end{equation}
where $\Omega$ is some antisymmetric matrix, with $\Omega\Omega ^*=-1$.
This is e.g.\ the way to have real spinors in $d=6$ Minkowski space.

Having two projections, to chiral spinors and to Majorana spinors,
one may ask whether they can be\textit{ combined, i.e.\ can we define a
reality condition respecting the chiral projection.} As
$(\Gamma_*)^*=(-)^{d/2+t}B\Gamma_* B^{-1}$ the condition is that
$d/2+t=0\ \mod 2$, i.e.
\begin{equation}
\mbox{MW spinor :}\ s-t= 0\ \mod 4\,.
\end{equation}
Table~\ref{tbl:MWSspinors} gives the resulting possibilities, and its caption
gives a summary.
\begin{table}[ht]
\begin{center}\begin{tabular}{|c|lr|lr|lr|lr|}\hline
d $\backslash$ t &\multicolumn{2}{c|}{0} &\multicolumn{2}{c|}{ 1}
& \multicolumn{2}{c|}{2} & \multicolumn{2}{c|}{3} \\ \hline
1 & M     & 1 & M    & 1 &       &   & &   \\
2 & M$^-$ & 2 & MW   & 1 & M$^+$ & 2 & &   \\
3 &       & 4 & M    & 2 & M     & 2 & & 4 \\
4 & SMW   & 4 & M$^+$& 4 & MW    & 2 & M$^-$ & 4 \\
5 &       & 8 &      & 8 & M     & 4 & M     & 4 \\
6 & M$^+$ & 8 & SMW  & 8 &  M$^-$& 8 & MW    & 4 \\
7 & M     & 8 &      & 16 &      & 16 & M    & 8 \\
8 & MW    & 8 & M$^-$& 16 & SMW  & 16 & M$^+$& 16 \\
9 & M     & 16 & M   & 16 &      & 32 &      & 32 \\
10& M$^-$ & 32& MW   & 16 & M$^+$ &32 & SMW  & 32 \\
11&       & 64& M    & 32 & M     & 32 & & 64 \\
12& SMW   & 64 & M$^+$& 64 & MW    & 32 & M$^-$ & 64\\ \hline
\end{tabular}\end{center}\caption{\sl Possible spinors in various dimensions,
and for various number of time directions (modulo 4). $M$ stands for Majorana
spinors. For even dimensions, $M^\pm$ indicates which sign of $\eta $
should be used. MW indicates the
possibility of Majorana--Weyl spinors. For even dimensions one can always have
Weyl spinors. Symplectic Majorana spinors are always possible when the
Majorana condition is not possible. SMW indicates the possibility of
symplectic Majorana--Weyl spinors. The number indicates the real dimension of
the minimal spinor. }\label{tbl:MWSspinors}\end{table}
\subsection{4 dimensions}
As an example take the Minkowski space in 4
dimensions. In order to have Majorana spinors we choose $\eta =1$,
and hence $\epsilon =1$.
There exists a Majorana representation
where the matrices $\gamma_\mu$ are real:
\begin{eqnarray}
&&\gamma_0=\pmatrix{0&\sigma_3\cr -\sigma_3&0\cr}\,,\qquad
\gamma_1=\pmatrix{0&-\unity _2 \cr -\unity _2 & 0\cr}\,, \qquad
\gamma_2=\pmatrix{0&-\rmi\sigma_2\cr \rmi\sigma_2 & 0\cr}\,, \nonumber\\
&&\gamma_3=\pmatrix{\unity _2 &0\cr 0 & -\unity _2\cr}\,,\qquad
\gamma_5=\pmatrix{0&-\rmi\sigma_1\cr \rmi\sigma_1 & 0\cr}\,.
\end{eqnarray}
The charge conjugation matrix is proportional to $\gamma _0$, say
${\cal C}=z\gamma_0$, with $|z|=1$.
As $A=\gamma_0$, the Majorana condition is
$\lambda^*=z\alpha\lambda$. Thus with an appropriate choice of signs,
$z\alpha=1$, Majorana spinors are just pure
real spinors in this basis.

In another representation that I present now, $\gamma _5$ is diagonal.
 That basis is
useful to have manifestly chiral spinors. These are then just spinors
that have only two upper components $\lambda^A$, while antichiral
have two lower components, denoted as $\lambda_{\dot A}$,
(where $A$ and $\dot A$ can be 1 or 2). The explicit realization is
\begin{equation}
\gamma_0=\pmatrix{0&\rmi\unity _2\cr \rmi\unity _2 & 0\cr}\,,\qquad
\gamma_i=\pmatrix{0&-\rmi\sigma_i\cr \rmi\sigma_i & 0\cr}\,, \qquad
\gamma_5=\pmatrix{\unity _2&0\cr 0& -\unity _2 \cr}\,,
\end{equation}
and with this choice
\begin{equation}
  {\cal C}=\pmatrix{\varepsilon& 0\cr 0&-\varepsilon}\,,\qquad  \mbox{with}
  \qquad \varepsilon =   \pmatrix{0&1\cr -1&0}\,.
\label{calCin4rep2}
\end{equation}
This basis makes the connection to the `2-component formalism' easy.
In that formulation, indices are raised or lowered with
$\varepsilon$ in a NW--SE convention
\begin{equation}
\chi^A=\varepsilon^{AB}\chi_B\,,\qquad \chi_A= \chi^B\varepsilon_{BA}\,,
\end{equation}
which implies $\varepsilon^{AB}\varepsilon_{BC}=-\delta^A{}_C$, and
the Majorana condition for a spinor $(\zeta^A, \zeta_{\dot
A})$  is then
\begin{equation}
\zeta_A=\left(  \zeta_{\dot A}\right)^*\,.
\end{equation}
I do not use this convention, but  use chiral spinors, avoiding the
indices.

A third representation that is often used, has diagonal $\gamma _0$.
\begin{equation}
\gamma_0=\pmatrix{\rmi\unity _2 &0\cr 0 & -\rmi\unity _2\cr}\,,\qquad
\gamma_i=\pmatrix{0&-\rmi\sigma_i\cr \rmi\sigma_i & 0\cr}\,, \qquad
\gamma_5=\pmatrix{0&-\unity _2\cr -\unity _2 & 0\cr}\,.
\label{explgamma}
\end{equation}
The charge conjugation is $ {\cal C}=\gamma _0\gamma _2$ and
the Majorana condition amounts to
\begin{equation}
  \lambda ^*=\pmatrix{0&-\sigma _2\cr \sigma _2&0}\lambda \,.
\label{Majoranaexpl}
\end{equation}

\subsection{Technical tips}\label{ss:remarksCl}
\subsubsection{Complex conjugation}\label{ss:scc}
To consider complex conjugation in practice, it is useful to consider
the operation $C$, defined by
\begin{equation}
  \lambda ^C\equiv \alpha ^{-1}B^{-1}\lambda ^*\,.
\label{defCconj}
\end{equation}
This is thus chosen such that the Majorana spinors are those for
which $\lambda ^C=\lambda $.
Note that the bar operation, defined by the first equation of
(\ref{Majorana}), on $\lambda ^C$ gives $\bar \lambda ^C$ as defined
in the second equation of (\ref{Majorana}). The square of this
operation gives the identity $(\lambda ^C)^C=\lambda $ if the right
hand side of (\ref{B*B}) is~1. In case that this is $-1$, we should
modify the definition of the $C$ operation similar to (\ref{symplMaj}).
One can check, using (\ref{AT}), that
\begin{equation}
  (\bar \lambda) ^*=\alpha \eta ^t \bar \lambda ^CB^{-1}\,.
\label{barlambda*}
\end{equation}

We are now equipped to discuss complex conjugation of bi-spinors.
First of all, we still have to determine whether complex conjugation
changes the order of fermions or not. In fact, both types of complex
conjugation are possible. I introduce a new sign factor to reflect
this choice, say $\beta $. Thus for two fermionic quantities,
\begin{equation}
  (\lambda \chi )^*=\beta\lambda ^*\chi ^*=- \beta \chi ^*\lambda ^*
  \,,\qquad \beta =\pm 1\,.
\label{defbeta}
\end{equation}
Consider now the complex conjugation of $\bar \chi M\lambda $, where
$M$ is some matrix in spinor space. With the previous equations one
easily finds
\begin{equation}
  \left(\bar \chi M\lambda\right) ^*=\beta \alpha ^2\eta ^t\bar \chi
  ^C B^{-1}M^*B\lambda ^C\,.
\label{ccbispinor}
\end{equation}
We can thus choose in any case $\alpha $ appropriate in order that
e.g.\ $\bar \chi \lambda $ is real.
Then we define $M^C\equiv B^{-1}M^*B$, thus e.g.
\begin{equation}
 \Gamma^C_a= -\eta (-)^t \Gamma _a\,\qquad
 \Gamma _*^C=(-)^{d/2+t}\Gamma _*\,,
\label{GammaC}
\end{equation}
the latter for even dimensions.

\noindent \emph{Rule: to perform complex conjugation of a bispinor,
first add a factor $\beta \alpha ^2\eta ^t$ and then replace fields and matrices
by their $C$-conjugates, for which (\ref{GammaC}) gives the important relations.}
\label{conclcc}

As an example consider Minkowski spaces where $\eta =\epsilon =1$,
thus in dimensions $d=2,3,4$ mod 8, containing
4, 10 and 11 dimensions. We choose $\alpha $ according to the
convention on complex conjugation of bi-spinors as $\beta \alpha
^2=1$. Then (\ref{ccbispinor}) implies that we can perform complex
conjugation by using the $C$ operation. Majorana spinors are thus
considered as just real, and the same applies for the ordinary gamma
matrices. However $\gamma _5$, the $\Gamma _*$ in 4 dimensions,
behaves as a pure imaginary matrix. In 10 dimensions this matrix can
also be considered as real, which reflects the possibility of
Majorana--Weyl spinors in that dimension.

\noindent \textit{Exercise:}

See how this should be set up in the case of symplectic Majorana
spinors. \QED
In extended supergravity in 4 dimensions, the chirality is
indicated by the position of the $i,j$ index
(index running over $1,\ldots N$ for $N$-extended
supergravity). See e.g.\ \eqn{chiralQ}.
 The choice of chirality for the spinor with an upper
(lower) index can change for each spinor. It is chosen conveniently
on the first occurrence of the spinor. With the rules of the $C$
operation, we see that h.c. effectively
interchanges upper and lower indices. The chiral spinor $Q^i$ is not
a Majorana spinor. Check that $(Q^i)^C=Q_i$.

\subsubsection{Gamma matrix manipulations}
I still give some useful identities for calculations in arbitrary
 dimensions \cite{JWvHAVP}.
For a product of two antisymmetrized gamma matrices, one can use
\begin{eqnarray}
&&  \Gamma _{a_1\ldots a_i}\Gamma ^{b_1\ldots b_j}=\sum_{k=|i-j|}^{i+j}
  \frac{i!j!}{s!t!u!}\delta ^{[b_1}_{[a_i}\cdots \delta ^{b_s}_{a_{t+1}}
  \Gamma _{a_1\ldots a_t]}{}^{b_{s+1}\ldots b_{j]}}
 \label{GammaGamma} \\
&& s=\ft12(i+j-k)\,,\qquad t=\ft12(i-j+k)\,,\qquad u=\ft12(-i+j+k)\nonumber\,.
\end{eqnarray}
In \cite{JWvHAVP} a few extra rules are given and a diagrammatic
technique is explained that is based on the work of Kennedy \cite{Kennedy}.

For contractions of repeated gamma matrices, one has the
formula
\begin{eqnarray}
  &&\Gamma_{b_1\ldots b_k} \Gamma _{a_1\ldots a_\ell }\Gamma^{b_1\ldots b_k}
  =c_{k,\ell }\Gamma _{a_1\ldots a_\ell }
  \nonumber\\
  &&c_{k,\ell }= (-)^{k(k-1)/2}k!(-)^{k\ell }\sum_i^{\min (k,\ell )}
  {\ell \choose i}{D-\ell \choose k-i}(-)^i\,,
\label{sandwich}
\end{eqnarray}
for which tables were given in \cite{JWvHAVP} in dimensions 4, 10, 11 and 12,
and which can be easily obtained from a computer programme.

Further, there is the Fierz relation. We know that
the gamma matrices are matrices in dimension $\Delta =2^{{\rm Int
~}d/2}$, and that a basis of $\Delta
\times \Delta $ matrices is given by the set
\begin{equation}
\left\{\unity ,\Gamma_a, \Gamma_{a_1a_2}, \ldots ,\Gamma ^{a_1\ldots a_{[D]}}
 \right\}
\qquad \mbox{where}\qquad
  \left\{ \begin{array}{ll}
     [D]=D & \mbox{for even }D \\{}
   [D]=(D-1)/2 & \mbox{for odd }D \,,
  \end{array}\right.
\label{basisFierz}
\end{equation}
of which only the first has nonzero trace. This is the basis of the
general Fierz formula for an arbitrary matrix $M$ in spinor space:
\begin{equation}
 M_\alpha {}^\beta\Delta= \sum_{k=0}^{[D]}(-)^{k(k-1)/2}\ft1{k!}
 \left(  \Gamma _{a_1\ldots a_k}\right)  _\alpha {}^\beta \,\,\trace
  \left( \Gamma ^{a_1\ldots a_k}M\right)\,.
  \label{Fierz}
\end{equation}
More useful Fierz identities can be found in \cite{Kennedy}.
\subsubsection{Spinor indices}
In these notes, I mostly omit spinor indices. However, in some
applications, this is difficult to accomplish or one prefers spinor
indices. I will give here a way to include spinor indices in a
practical way, independent of the dimension of spacetime.
First of all, spinors get a lower spinor index. So for
a spinor $\lambda $, I write $\lambda _\alpha $. Gamma matrices act
on these spinors, and therefore the expression $\Gamma_a\lambda $
becomes $(\Gamma _a)_\alpha {}^\beta \lambda _\beta $. Note the
position of the spinor indices on a usual Gamma matrix. This is then
the same for products of Gamma matrices.

Then I want to be able to raise and lower indices. Indeed, e.g.\ to
discuss symmetries of Gamma matrices, the $(\Gamma _a)_\alpha {}^\beta
$ can not be used, as we can not interchange upper and lower indices.
To that purpose, I introduce matrices ${\cal C}^{\alpha \beta }$ and
${\cal C}_{\alpha \beta }$, which will be related to the charge
conjugation matrix below.
The convention that I adopt, is that I raise and lower indices always
in the NorthWest--SouthEast (NW-SE) convention. That means that the
contraction indices should appear in that relative position, i.e.\
\begin{equation}
  \lambda ^\alpha ={\cal C}^{\alpha \beta }\lambda _\beta \,,\qquad
  \lambda _\alpha = \lambda ^\beta {\cal C}_{\beta \alpha }\,.
\label{NWSE}
\end{equation}
In order for these two equations to be consistent, we should have
\begin{equation}
  {\cal C}^{\alpha \beta }{\cal C}_{\gamma \beta  }=\delta ^\alpha _\gamma
  \,,\qquad {\cal C}_{ \beta \alpha  }{\cal C}^{ \beta\gamma }=\delta
  ^\gamma _\alpha\,.
\label{CCcontr}
\end{equation}
I choose the identifications such that the Majorana conjugate
$ \bar \lambda $ is written as $\lambda ^\alpha $. Comparing
(\ref{NWSE}) and (\ref{Majorana}), we conclude that
$ {\cal C}^{\alpha \beta }$ is ${\cal C}^T$, and $ {\cal C}_{\alpha \beta
}$ is then $ {\cal C}^{-1}$. One may check that the symmetries of
$ {\cal C}\Gamma ^{(n)}$ or $\Gamma ^{(n)}{\cal C}^{-1}$, which were
discussed before, is now the symmetry of $ (\Gamma _a)^{\alpha \beta
}$ or $(\Gamma _a)_{\alpha \beta }$.

\noindent \textit{Exercise :} Check that $\bar \chi \lambda =
\chi ^\alpha \lambda _\alpha =\pm\chi _\alpha \lambda^ \alpha$ where
the sign is $-\epsilon $, i.e.\ $+(-)$ if the charge
conjugation matrix is (anti)symmetric. The same is true for spinor
indices at any place: $(\Gamma _a)_\alpha {}^\beta\lambda _\beta =
\pm (\Gamma _a)_{\alpha \beta }\lambda ^\beta $.


\section{Supersymmetry algebras}\label{s:susyalg}
\subsection{Haag--{\L}opusza\'nsky--Sohnius result as introduction}
  \label{ss:HLS}
After the introduction with bosonic symmetries, and the recapitulation
of spinor properties, I can finally
consider fermionic symmetries. This means that the parameters of
some transformations can be anticommuting numbers. For the generators
we then have anticommutators rather than commutators. Indeed,
consider
\begin{equation}
\delta(\epsilon)=\epsilon^A Q_A\,,
\end{equation}
for anticommuting parameters $\epsilon^A$ and hence operators $Q_A$,
which I will denote further as `odd', distinguishing them from the
bosonic `even' operators.
We have then
\begin{equation}
[\delta(\epsilon_1),\delta(\epsilon_2)]=\epsilon_2^B\epsilon_1^A
(Q_AQ_B+Q_BQ_A)\equiv \epsilon_2^B\epsilon_1^A \left\{Q_A, Q_B\right\}\,.
\end{equation}
Here, I introduced a notation for the anticommutator, and I
also introduce a general notation
\begin{equation}
[T_A, T_B\} = T_A T_B -(-)^{AB} T_B T_A\,.
\label{defcac}
\end{equation}
The notation $[,\}$ is a commutator except when both operators are
odd, when it is an anticommutator.
In the notation $(-)^A$ one should thus understand
$A$ as an even number, e.g.\ 0, if $T_A$ is even, and $A$ as an
odd number, e.g.\ 1, if $T_A$ is odd.

These algebras should also satisfy a Jacobi identity. Keeping the
parameters in place, such a Jacobi identity takes the same form as the
pure bosonic one:
\begin{equation}
\left[\epsilon_1^A T_A,\left[\epsilon_2^B T_B, \epsilon_3^C
T_C\right]\right] + \mbox{ cyclic in 1, 2, 3 }=0\,.
\end{equation}
Removing the parameters, signs appear, but an easy form to remember
is
\begin{equation}
\left[ \left[T_A , T_B\right\}, T_C\right\}=
\left[ T_A ,\left[ T_B, T_C\right\}\right\}-(-)^{AB}
\left[ T_B ,\left[ T_A , T_C\right\}\right\}\,.  \label{Jacobieasy}
\end{equation}
The last sign factor just reflects the symmetry of the commutator in
the left-hand side.

These definitions lead to a lot of extensions of algebras. We will come
back to the mathematical part in
section~\ref{ss:superalgebras}. First we consider the physical
input. As a generalisation of the result of \cite{ColemanMandula},
reviewed in section~\ref{ss:ColemanMandula},
Haag, {\L}opusza\'nsky and Sohnius \cite{HLS} considered the same
question allowing for fermionic symmetries. As for \cite{ColemanMandula},
 their result
contained two parts: one general possibility (which we will label as the
super-\Poin\ algebra), and one for massless fields only, the superconformal
algebras. An important part of their result is that super-\Poin\
algebras may contain `central charges', which we will discuss in
section~\ref{ss:cc}.
They did suppose a flat space, but we are also interested
in the possibility of a curved space, i.e.\
the super-(anti) de Sitter algebras.

First I will give some qualitative remarks on supersymmetry in
general. Then I will consider the super-\Poin algebras, first without
central charges in section~\ref{ss:susyalgebras}, then with
 central charges in section~\ref{ss:cc}. I
will turn to super-anti-de Sitter algebras in section~\ref{ss:sadS},
but will have to interrupt that analysis to
give an overview of simple superalgebras in general
(section~\ref{ss:superalgebras}). Finally I will turn to the
superconformal algebras in section~\ref{ss:scalg}.
\subsection{Basic properties of \Poin\ supersymmetry}
In general the algebra contains
respectively the \Poin, conformal or (anti) de Sitter algebra, and a number of
(Lorentz) scalar generators on the even side, and spin-1/2 odd generators.
For the \Poin\ case these are generators $Q_\alpha^i$, where $\alpha$ is
a spinor index, and $i$ is a representation index for some part of the bosonic
scalar generators outside of the \Poin\ algebra. Remember that the whole
discussion here is in 4 dimensions, where the spinors can be split in a left
and a right chirality, which we indicate by the position of the $i$ index:
\begin{equation}
Q^i_\alpha =\ft12 (1-\gamma_5)_\alpha {}^\beta  Q^i_\beta \,,\qquad
Q_{i\alpha }=\ft12 (1+\gamma_5)_\alpha {}^\beta Q_{i\beta }\,.\label{chiralQ}
\end{equation}
Their anticommutator takes the form
\begin{equation}
\{ Q_\alpha^i, Q_{\beta j}\}=\left( \gamma^\mu {\cal C}^{-1}\right)
_{\alpha\beta}P_\mu \delta^i_j\,, \label{susyalgebra}
\end{equation}
and I will return to the meaning of the charge conjugation matrix, {\cal
C}, in section~\ref{ss:spinors}.
This is the defining relation of supersymmetries: they square to the
translations. One of the consequences is that it has a dimension
$1/2$, where the dimension of translations is conventionally taken to
be~1, i.e.\ as a mass dimension. So e.g.\ if a scalar $\phi$ of dimension~1
transforms in a fermion\footnote{Spinor indices are deleted, and details
of the notation will follow later.}
\begin{equation}
\delta_Q \phi= \bar \epsilon \lambda\,,
\end{equation}
then, as $\epsilon$ has dimension $-1/2$, the dimension of $\lambda$ is $1/2$
higher than that of $\phi$, and its supersymmetry transformation contains
$\phi$ always with a derivative or a mass. Of course, in minimal actions of
scalars and spinors, the former have dimension $1$, and the latter $3/2$, so we
are not unhappy with this conclusion.

\noindent \textbf{Theorem:} \emph{There are an equal number of bosonic and fermionic
degrees of freedom in any realization of the supersymmetry algebra
when translations are an invertible operation.}
\par
\noindent Consider the commutator \eqn{susyalgebra} on the space of all bosons.
The first $Q$ transforms the space to a space of fermions, while the second one
brings us back to bosons, translated by $P_\mu$.
\[
\mbox{bosons}\stackrel{Q}{\longrightarrow }
\mbox{fermions}\stackrel{Q}{\longrightarrow }
\mbox{bosons translated by }  P_\mu
\]
The latter is an invertible operator, which proves that the last
space is as large as the first one, and thus also the middle one, the
space of the fermions has the same dimension. This proves the
well-known fact that there are an equal number of bosonic and of
fermionic states. It should hold when the algebra \eqn{susyalgebra}
holds. So this equality of bosonic and fermionic states should hold
e.g.\ for on-shell states, but also for off-shell states if the
algebra is also realized off-shell. On the other hand, it is not a
general fact of a superalgebra. One needs the invertibility of the
square of fermionic generators. This does not hold for massless
states in an Euclidean spacetime, or in 1 dimension. In these cases
the $P^2=0$ condition for a massless state implies $P=0$, and there is
thus no invertibility.

\subsection{Super-\Poin\ algebras}\label{ss:susyalgebras}
Now we can study the superalgebras which are important in
supersymmetry. I start with the
result of \cite{HLS} in 4 dimensions. Apart from the bosonic \Poin\
algebra \eqn{Poinalgebra}, and the defining relation of
supersymmetry \eqn{susyalgebra}, there is of course the statement
that the supersymmetries are spinors under the Lorentz group\footnote{
As the reader probably already noticed, we denote $\Gamma$ matrices
in 4 dimensions as $\gamma_\mu$}:
\begin{equation}
\ [ M_{\mu\nu} , Q_{\alpha}^i \ ] = -\ft14
(\gamma_{\mu\nu})_{\alpha}{}^\beta Q^i_\beta    \,.
\end{equation}
The supersymmetries may also rotate under an automorphism group
$T_A$:
\begin{equation}
\ [ T_A , Q_{\alpha}^i \ ] =  (U_A)^i{}_j Q_{\alpha}^j   \,,\qquad
\ [ T_A , Q_{\alpha i} \ ] =  (U_A)_i{}^j Q_{\alpha j}\,.
\end{equation}
The latter equation is obtained from the former by complex
conjugation, thus the complex conjugate of $(U_A)^i{}_j$ is denoted
as $(U_A)_i{}^j $, consistent with the rule mentioned at the end of
section~\ref{ss:scc}. The Jacobi identity $[TTQ]$ implies that
these matrices form a representation of the algebra of the $T_A$. And
as $T_A$ has to commute with $P_\mu$, as follows already from the
Coleman--Mandula result, it is easy to check that the $[TQ_iQ^j]$
Jacobi identity implies that
\begin{equation}
(U_A)_i{}^j =-(U_A)^j{}_i\equiv -\left((U_A)_j{}^i \right) ^*\,.
\end{equation}
Thus these are unitary matrices and
 the automorphism group is $U(N)$, for $i=1,\ldots ,N$.

We can make a general statement about the isometry groups in view of
table~\ref{tbl:MWSspinors}. The 4-dimensional $t=1$ example is
general for even dimensions where no Majorana--Weyl spinor is
possible, thus indicated by $M$ in the table. Then the automorphism
group is $U(N)$.  In fact, in this formulation we have not used the
Majorana property, but rather the Weyl-spinor formulation. In all even
dimensions we can use Weyl spinors, and this leads to spinors of the
same dimension as the Majorana ones (they are the same in another notation).
If we would have the
possibility of Majorana--Weyl, then there is an extra reality
condition, and we have two factors as automorphism groups of the left
and right spinors, i.e.\ we obtain $SO(N_L)\times SO(N_R)$. For odd
dimensions, we can not use the chiral formulation, and if there are
Majorana spinors, the result is $SO(N)$. If there are symplectic
Majorana conditions, then the same argument of the Jacobi identity
$[TQQ]$ leads to the preservation of the symplectic metric, and the
automorphism algebra is thus reduced to $USp(N)$ (even
$N$)\footnote{There is often confusion about the notation of $Sp$
groups. We use notation such that $Sp(2)$ is the smallest symplectic
group, see section~\ref{ss:superalgebras}.}.  If there is
a SMW spinor, then there are two such factors. In summary, the
automorphism groups are obtained from the entries of
table~\ref{tbl:MWSspinors}:
\begin{eqnarray}
M\mbox{ and }d\mbox{ odd}\,\ &:& SO(N)\nonumber\\
M\mbox{ and }d\mbox{ even}\ &:& U(N)\nonumber\\
MW\ \ \ \ \ \ \ \ \ \ \ \ &:& SO(N_L)\times SO(N_R) \nonumber\\
S \mbox{ (empty in table)}&:& USp(N)  \nonumber\\
SMW\ \ \ \ \ \ \ \ \ \ \, &:& USp(N_L)\times USp(N_R)\,.
\end{eqnarray}

We can now write a list of supersymmetry theories in
Min\-kowski spaces of various dimensions and with various extensions.
Table~\ref{tbl:sgdimext},
\begin{table}[ht]
\begin{center}\begin{tabular}{||l|cccc||}\hline
d $\backslash$ Q &32& 16  &8  & 4  \\ \hline
11& X & &&  \\[2mm]
10& $\stackrel{ IIB}{(2,0)}\ \stackrel{ IIA}{(1,1)}$ &
$\stackrel{I}{(1,0)}$
& & \\[2mm]
9 & X & X &&\\[2mm]
8 & X & X &&\\[2mm]
7 & X & X &&\\[2mm]
6 & $(2,2)$& $(2,0) \ (1,1)$ &
$(1,0)$ &\\[2mm]
5 & X & X & X &\\[2mm]
4 &$N=8$ &$N=4$ & $N=2$
&$N=1$\\  \hline
\end{tabular}\end{center}
\caption{\sl Pure supergravity theories in dimensions $4\leq
d\leq11$ with the number of independent
supercharges equal to $Q=32$, 16, 8 and 4. In 3 spacetime
dimensions, pure supergravity does not describe propagating
degrees of freedom and is a topological theory.}\label{tbl:sgdimext}
\end{table}
which I essentially copied from \cite{BernardJan}, gives an overview
of possible supersymmetry models\footnote{Note that the theories are
not always unique. Certainly if one allows arbitrary signs in kinetic
terms for some fields.}. This follows mainly from
table~\ref{tbl:MWSspinors}. The limit of 32 supercharges comes from
the requirement that one can not construct field theories with
helicities larger than 2. I have not explained representations
of the algebras, and therefore can not really prove this. However,
one can understand that in 4 dimensions any supersymmetry raises or
lowers the helicity by $1/2$. Therefore with $N=8$ we already fill
the whole range from 2 to $-2$. Similarly rigid supersymmetry models
are only possible for $N\leq 4$, i.e.\ $Q\leq 16$.  We still remark that other values
of $Q$, i.e.\ $Q=12,20,24$, are also possible in $d=4$, but are rarely
used.

Let me summarize  the result for the super \Poin\ algebras in 4
dimensions, omitting the automorphisms (as can consistently be done
for the super-\Poin\ case):
\begin{eqnarray}
[M_{\mu\nu} , M^{\rho\sigma}]&=&
-2\delta_{[\mu}^{[\rho} M_{\nu]}{}^{\sigma]} \,, \nonumber\\
\ [P_{\mu} , M_{\nu\rho}\ ] &=&\eta_{\mu[\nu} P_{\rho]}\,, \nonumber\\
\ [P_{\mu} , P_{\nu}\ ] &=&  0 \nonumber\\
\{ Q_\alpha^i, Q_{\beta j}\}&=&\left( \gamma^\mu {\cal C}^{-1}\right)
_{\alpha\beta}P_\mu \delta^i_j\nonumber\\
\ [ M_{\mu\nu} , Q_{\alpha}^i \ ]& =& -{1\over 4}
(\gamma_{\mu\nu}Q)_{\alpha} ^i \nonumber\\
\ [ P_{\mu} , Q_{\alpha}^i \ ]& =&0\,.
\end{eqnarray}
Remark that we may also consistently remove the Lorentz algebra
generators $M$. Thus the algebra is a semi direct sum of the part
with $P$ and $Q$ only, with the Lorentz algebra and with the
automorphism algebra.
\par
Let us check the reality property of the anticommutator of 2
supersymmetries, and generalize to other dimensions and
signatures. Complex conjugation gives
\begin{equation}
\beta \alpha ^2 B\left\{ (Q^i)^C, (Q_j)^C\right\} B^T =
 -\eta (-)^t B\Gamma_\mu  C^{-1}B^T \eta ^t P^\mu \,,
\label{ccQQ1}
\end{equation}
where spinor indices are implicit. The factor $\beta $ results from
the fact that complex conjugation may changes the order of fermions
if $\beta =-1$. Note that it does not change the order in which the
generators act. Thus there is only that factor, but the operators
keep their order. The $\beta $ spinor index, which was explicit in
(\ref{susyalgebra}), is at the right side of the equation, and that
is the reason of $B^T$ at the end of the left hand side, as $Q^*_\beta
= \alpha Q^C_\gamma (B^T)^\gamma {}_\beta $. For the right hand side
of (\ref{ccQQ1}), I used
\begin{equation}
  {\cal C}^{-1, *}={\cal C}^T=\eta ^t B{\cal C}^{-1}B^T\,,
\label{Cauxeqn}
\end{equation}
and the bosonic operators are real. Remark that if we move the $\eta
^t$ to the left hand side, we find here again the factors of
(\ref{ccbispinor}) and (\ref{GammaC}). Such commutators in general
behave as bi-spinors, as one can imagine multiplying them with spinor
parameters at the left and right. Thus they follow the rule
mentioned at page~\pageref{conclcc}.

The symmetry of the commutator
(and of $\Gamma _\mu {\cal C}^{-1}$) implies that in the right hand
side in 4 dimensions we can use
\begin{equation}
  \left\{ (Q^i)^C, (Q_j)^C\right\}= \left\{Q_i, Q^j\right\}=
  \left\{Q^j,Q_i\right\}\,.
\label{symQQC}
\end{equation}
Then the reality consistency is proven if we take (for
Minkowski spaces with $\eta =1$) conventions with
\begin{equation}
  \beta \alpha ^2=1\,.
\label{betaalpha2}
\end{equation}
\par
With these methods it is easy to check how this can be generalised to
other dimensions, with other types of spinors.
E.g.\ in $d=11$ (Minkowski, $t=1$) there are no chiral
spinors, and we just have
\begin{equation}
\{ Q_\alpha, Q_{\beta}\}=\left( \gamma^\mu {\cal C}^{-1}\right)
_{\alpha\beta}P_\mu \,.
\end{equation}
The reality goes as above, one can just neglect the $i$ indices.

In $d=10$  one can have Majorana--Weyl spinors.
For having a similar $\{Q,Q\}=P$ anticommutator, the $\Gamma _\mu
P^\mu$ should thus appear in the anticommutator between two
supersymmetries of the same chirality.
One can ask in general the question whether one can have
the translations $P_\mu $ in the right hand side of the
anticommutator between two chiral supersymmetries $\{ Q_L, Q_L\}$.
If that is the case, then
\begin{eqnarray*}
\{ Q_L, Q_L\}&=&\ft12 \left( 1+\Gamma _*\right)
\gamma^\mu {\cal C}^{-1} P_\mu \ft12 \left( 1+\Gamma _*\right)^T
  \nonumber\\
&=&\ft14\left( 1+\Gamma _*\right)
 \left( 1-(-)^{d/2}\Gamma _*\right)\gamma^\mu  P_\mu{\cal C}^{-1}
 \nonumber\\
 &=&\ft14 \left(1-(-)^{d/2}\right)\left( 1+\Gamma _*\right)
\gamma^\mu {\cal C}^{-1} P_\mu\,,
\label{QLQLcons}
\end{eqnarray*}
where I used that
\begin{eqnarray}
\Gamma_*^T&=&(-\rmi)^{d/2+t}\Gamma_d^T\ldots \Gamma_1^T
=(-\rmi)^{d/2+t} {\cal C}\Gamma_d\ldots
\Gamma_1{\cal C}^{-1}\nonumber\\
&=&(-)^{d/2}{\cal C} \Gamma_* {\cal C}^{-1}\,.
\end{eqnarray}
Thus $d/2$ should be odd, as it is the case in 10 dimensions.
If there is one such chiral supersymmetry, we denote this as
 $(1,0)$ supersymmetry. If there are two, then this is
$(2,0)$  (which is usually called IIB). There can be two
supersymmetries of opposite chirality which is denoted as
 $(1,1)$, or IIA supersymmetry. Remark that the result
 (\ref{QLQLcons}) also checks that in 4 dimensions the $P_\mu $ does
 not appear in the anticommutator of 2 generators of the same
 chirality, but, as immediately shown in (\ref{susyalgebra}), in the
 anticommutator of a
chiral and an antichiral supersymmetry.
\par
\noindent \textit{Exercise.} Check that for $d/2$ even, consistency is
obtained if $P_\mu $ appears in the anticommutator of a chiral with
an antichiral supersymmetry. Further check that for $d/2$ odd, the
anticommutator of two supersymmetries of the same chirality can only
have $\Gamma ^{(n)}{\cal C}^{-1}$ for $n$ odd, while for opposite
chirality $n$ should be even. This is reversed for $d/2$ even.
\QED
An important consequence of supersymmetry, which we did not mention
yet, is contained in the {\em positivity statements} of the energy.
To consider these, we want the anticommutator between a
supersymmetry and its complex conjugate:
\begin{equation}
  \left\{ Q_i, (Q_j)^*\right\} =\alpha \left\{ Q_i,
  (Q_j)^C\right\}B^T= \alpha \delta _i^j \Gamma _\mu A^{-1}P^\mu \,,
\label{QQ*}
\end{equation}
where the $i,j$ indices are put as for $d=4$, but one can forget them
for e.g.\ $d=11$. The left hand side is real and positive definite if
we do not interchange spinors in complex conjugation ($\beta =1$).
Then we can choose $\alpha =1$, consistent
with (\ref{betaalpha2}). For $\beta =-1$ the left hand side is
imaginary, but $\alpha $ is then an imaginary unit. We can therefore
conclude that we can choose $\alpha $, consistent
with (\ref{betaalpha2}), such that
\begin{equation}
  \alpha ^{-1}\left\{ Q_i, (Q_j)^*\right\}\geq 0\,.
\label{posQQ}
\end{equation}
As $A=\Gamma _0$ in Minkowski spaces, this exhibits the positive
energy statements in supersymmetry. For a state at rest, i.e. $P^\mu
=\delta ^\mu _0 M$, this implies $M\geq 0$.

\subsection{\Poin\ algebras with central charges} \label{ss:cc}
Finally we come to an extra possibility, which are the central
charges. They have been found in the classification of
Haag--{\L}opusza\'nsky--Sohnius \cite{HLS}, and were realized in the
massive (and short) hypermultiplets in \cite{Hypermultiplet}. This was
in 4-dimensional Minkowski space, to which we will restrict first,
 before discussing the generalization to higher dimensions.

I did not yet give the commutator of two supersymmetries of equal
chirality. As I do not want new symmetries with Lorentz indices, the
general possibility is
\begin{equation}
\{ Q_{\alpha i}, Q_{\beta j}\}={\cal C}^{-1}_{\alpha\beta}\omega_{ij}^M
Z_M\,,\label{QQZ}
\end{equation}
where $M$ labels different central charges $Z^M$, characterized by
independent antisymmetric matrices $\omega _{ij}^M$. Indeed,
the properties of spinors imply that $\omega_{ij}^M$ should be antisymmetric.
Now one has to check all Jacobi identities, and at the end one
obtains that these $Z^M$ should commute with all other generators
including the supersymmetries. Therefore, these are called `{\em
central charges}'. Considering finally $[TQQ]$, we find a
restriction on the automorphism group. Indeed, one obtains
\begin{equation}
U_i{}^k \omega_{kj}-U_j{}^k\omega_{ki}=0\,.
\end{equation}

Let us consider as an example $N=2$. Then for $\omega_{ij}$,
there is only $\varepsilon_{ij}$, as this is the only antisymmetric
tensor. We thus have 1 (complex) central charge:
\begin{equation}
\left\{ Q_{\alpha i}, Q_{\beta j}\right\}= {\cal
C}^{-1} _{\alpha\beta}\varepsilon_{ij} Z\,.   \label{Zalgebra}
\end{equation}
If there
is such a central charge, then the automorphism group is not $U(2)$,
but rather the $U(1)\times U(1)$ which commutes with $\varepsilon_{ij}$.

Central charges are essential in various application due to the
modified positivity statements. Indeed, the modified algebra allows
us to obtain another positivity condition.

I will show the argument for $N=2$, but it can easily be generalized.
The complex conjugate of (\ref{Zalgebra}) is\footnote{I use here
$Q^{*\alpha i}\equiv (Q_{i\alpha })^*$. The tensor $\varepsilon ^{ij}$
is the same as $\varepsilon _{ij}$, namely $\varepsilon ^{12}=\varepsilon
_{12}=1$, but I write one or the other according
to the index structure of the equation.}
\begin{equation}
\beta \left\{ Q^{* \alpha i}, Q^{* \beta j}\right\}=-\varepsilon^{ij}
 {\cal C} ^{\alpha\beta} Z^* \,.
\end{equation}
To diagonalize the set of anticommutators for the supersymmetries,
I define
\begin{equation}
A_{\alpha i}\equiv Q_{\alpha i}+\alpha ^{-1}e^{\rmi\theta}
\varepsilon_{ij}{\cal C}^{-1}_{\alpha\beta}
Q^{* \beta j}\,,
\end{equation}
where $e^{\rmi\theta}$ is an arbitrary phase factor.
 The complex conjugate is
\begin{equation}
A^{* \alpha i}\equiv \left( A_{ \alpha i}\right) ^*=Q^{* \alpha i}+
\alpha e^{-\rmi\theta}\varepsilon^{ij}Q_{j\beta}
{\cal C}^{\beta\alpha} =\alpha e^{-\rmi\theta} \varepsilon^{ij}A_{j\beta}
{\cal C}^{\beta\alpha}\,.
\end{equation}
Consider now their anticommutators on a state for which
$P^\mu =\delta^\mu_0 M$:
\begin{equation}
\alpha ^{-1}\left\{ A_{\alpha i},A^{* \beta j}\right\}= \delta_i^j
\delta_\alpha^\beta (2M+ Ze^{-\rmi\theta}+Z^* e^{\rmi\theta})\,.
\end{equation}
As the left hand side is a positive definite operator, the
right hand side should be positive for all $\theta$, which
shows that
\begin{equation}
  M\geq |Z|\,.
\label{BPSbound}
\end{equation}
This is an important result, giving a lower limit to the
mass, dependent on the central charge. This is called the {\em BPS bound}.
One can now also see that if the
equality is satisfied for a state (a `BPS state'), then that state is
invariant under part of the supersymmetry generators.
That state preserves thus part of
the supersymmetry. Another role of the central charges, is that one can
have supersymmetry multiplets with central charges such that the bound is
saturated. Such multiplets have less components than others, and are therefore
 called short multiplets.
 These are the characteristics of algebras with
central charges.

Knowing of this possibility of central charges in 4 dimensions,
 one may try to
generalize it to {\em higher dimensions}. One looks again for
generators in the $\{Q,Q\}$
anticommutator. This was first considered in the rheonomic approach
to $d=11$ supergravity \cite{d11rheo}, for solving $d=10$
super-Yang--Mills superspace constraints \cite{d10sYMssp}, and while
looking for superconformal algebras in higher dimensions
\cite{JWvHAVP}. The presence of these operators will lead to
positivity bounds, similar to the one that we found in $N=2$, $d=4$.

Consider as an example $d=11$, with one supersymmetry.
In the $\{Q,Q\}$ anticommutator one can think of all sorts of objects
\begin{equation}
\{ Q_\alpha, Q_\beta\}= \sum_{n} \left( \Gamma^{(n)}{\cal
C}^{-1}\right) _{\alpha\beta} Z_{(n)} \,.
\end{equation}
But of course, the right hand side should also be symmetric. Considering
table~\ref{tbl:epseta}, we see that this restricts us already to
$n=1,2 \ \mod 4$,
and for an odd dimension, the independent ones are those up to $n=(d-1)/2$.
This implies that we have at most
\begin{equation}
\{ Q_\alpha, Q_\beta\}=
\left( \Gamma^\mu{\cal C}^{-1}\right) _{\alpha\beta}P_\mu +
\left( \Gamma^{\mu\nu}{\cal C}^{-1}\right) _{\alpha\beta}Z^2_{\mu\nu} +
\left( \Gamma^{\mu\nu\rho\sigma\tau}{\cal C}^{-1}\right) _{\alpha\beta}
Z^5_{\mu\nu\rho\sigma\tau} \,. \label{algPoind11cc}
\end{equation}
The $Z^2$ and $Z^5$ are thus new generators in this anticommutator.
In fact, they are not `central' in the algebra, i.e.\ they do not
commute with all the other generators. As they have
spacetime indices, they do not commute with the Lorentz rotations.
The terminology which is in use, refers to the fact that these generators
have the same physical consequences as the real central charges in
4 dimensions, i.e.\ they lead to BPS bounds similar to (\ref{BPSbound}),
with the limit obtained by states which preserve part of the supersymmetry.

In 4 dimensions the scalar central charges are gauged by a vector,
and the charges are produced by a particle. In 11 dimensions various
solutions produce such charges, e.g.\ M2 or M5 branes, as
has be reviewed by J. Gomis in this school, and I refer the reader to
\cite{Malgebras,Mfromsa}.

\subsection{Super-anti-de Sitter algebras}\label{ss:sadS}
Remember that in the bosonic case, the \Poin\ algebra is not a simple
algebra, but a contraction $R\rightarrow \infty$ of the simple
(anti) de Sitter algebra \eqn{adSalgebra}. The same
will apply for the corresponding superalgebras. But in this section
we will first go step by step to construct a super anti-de Sitter
algebra.

Due to the last commutator of \eqn{adSalgebra}, we find that we need
a modification with respect to the super-\Poin\ algebra to satisfy the
Jacobi identity $[P,P,Q]$, see \eqn{Jacobieasy}. Indeed, it can not
be satisfied anymore with $[P,Q]=0$. For covariance (or consistency
with Lorentz transformations, i.e.\ $[M,P,Q]$ Jacobi), we should have
(at least for $N=1$)
\begin{equation}
[P_\mu, Q]=x\Gamma_\mu Q\,, \label{PQcom}
\end{equation}
where $x$ is an arbitrary number. But
taking the complex conjugate and using \eqn{realcond} and \eqn{Gammacc}
we obtain
\begin{equation}
x=-\eta(-)^t x^* \,.
\end{equation}
Continuing with the $[P,P,Q]$ Jacobi identity,
we get (using  \eqn{adSalgebra})
\begin{equation}
x^2=\frac{1}{16\,R^2}\,.   \label{x2adS}
\end{equation}
Specifying to $t=1$, we thus find that for $\eta=1$, which is e.g.\
the case in $d=11$ and $d=4$ if there are Majorana spinors, $x$
should be real. Thus $R^2$ is positive, and this implies, see the
discussion following \eqn{adSalgebra}, that we can only have the
anti-de Sitter algebra, rather than de Sitter. Indeed, the de Sitter
algebra would require $R^2<0$, which is not
consistent with \eqn{x2adS} and real $x$. For $N=1$ in $d=4$ or $d=11$
this thus proves that one should consider anti-de Sitter, rather than de Sitter
algebras. For extended supersymmetry,  one could have
introduced an antisymmetric matrix in the $ij$ indices in
\eqn{PQcom}, and my arguments for anti-de Sitter
are thus not complete. For $d=10$ with $N=1$ chiral supersymmetry one can not
have either de Sitter nor anti-de Sitter, as \eqn{PQcom} is not
consistent with the chirality projection.
\par
Continuing in this way with Jacobi identities \cite{JWvHAVP}, we
obtain the following result for $d=4$, $N=1$:
\begin{eqnarray}
 \left\{ Q_{L\alpha},\,Q_{R\beta}\right\}&=&
 \left(  \Gamma^\mu  {\cal C}^{-1} \right) _{\alpha\beta} P_\mu   +
 2x
 \left(  \Gamma^{\mu\nu}  {\cal C}^{-1} \right)
 _{\alpha\beta}M_{\mu\nu}\nonumber\\
{} [P_\mu,Q]&=&  x \Gamma_\mu Q  \nonumber\\ {}
[M_{\mu\nu},Q]&=&-\ft14 \Gamma_{\mu\nu}  Q \nonumber\\
\left[M_{\mu\nu},M_{\rho\sigma}\right]
&=&\eta_{\mu[\rho}M_{\sigma]\nu}
-\eta_{\nu[\rho}M_{\sigma]\mu}\nonumber\\ \left[P_\mu ,
M_{\nu\rho}\right]&=&\eta_{\mu[\nu}P_{\rho]}\nonumber\\ \left[
P_\mu,P_\nu\right]&=& 8\,x^2\,M_{\mu\nu}\,,\label{superadSd=4}
\end{eqnarray}
while in $d=11$ we obtain
\begin{eqnarray}
 \left\{ Q_\alpha,\,Q_\beta\right\}&=&
 \left(  \Gamma^\mu  {\cal C}^{-1} \right) _{\alpha\beta} P_\mu   +
 2x \left(  \Gamma^{\mu\nu}  {\cal C}^{-1} \right)
 _{\alpha\beta}M_{\mu\nu}\nonumber\\ &&
+\frac{1}{5!}\left( \Gamma^{\mu\nu\rho\sigma\tau}{\cal C}^{-1}\right) _{\alpha\beta}
Z^5_{\mu\nu\rho\sigma\tau} \nonumber\\
 \left[ Z^5_{\mu\nu\rho\sigma\tau},Q \right]&=&
 \frac{x}{2}\Gamma_{\mu\nu\rho\sigma\tau}Q \nonumber\\
\left[Z^5_{\mu_1\ldots \mu_5} , M^{\nu\rho}\right]&=&
5\delta_{[\mu_1}^{[\nu}Z^{5\,\rho]}{}_{\mu_2\ldots \mu_5]}\,,
\label{superadSd11}
\end{eqnarray}
with the other rules of \eqn{superadSd=4} unchanged.
Remark that we need now also the 5-index `central charge' generator in
order to be able to close the algebra. However, it is even less
`central' than before, because this generator now also has a non-zero
commutator with $Q$. The $Z^2$ of \eqn{algPoind11cc} has been
identified here with the Lorentz transformations.

In fact, the $d=11$ super adS algebra is not of the `HLS' form. Its
bosonic algebra is not
$adS\times G$. Here it is $Sp(32)$. Indeed the generators of $P_\mu$,
$M_{\mu\nu}$ and $Z^5$ constitute
\begin{equation}
{11 \choose 1}+{11\choose 2}+{11 \choose 5}=11+55+462=528=\frac{32 . 33}{2}
\end{equation}
generators. We can combine them in the $32\times 32$ matrix form
\begin{equation}
M = \frac{1}{x} \Gamma^\mu P_\mu + \Gamma^{\mu\nu}M_{\mu\nu}+
\frac{1}{x}  \Gamma^{\mu\nu\rho\sigma\tau}Z^5_{\mu\nu\rho\sigma\tau}
\end{equation}
to see this algebra explicitly. The full superalgebra is
$OSp(1|32)$, but for that we should first give an introduction to
superalgebras. Then we will be able to
classify also the super adS algebras in other dimensions.

\subsection{Superalgebras}\label{ss:superalgebras}

Lie superalgebras have been classified in \cite{LieSA}.
I do not have the time to go through the full classification mechanism
of course, but will consider the most important superalgebras, the
 `simple Lie superalgebras', which have no non-trivial
invariant subalgebra. However, one should know that in superalgebras
there are more subtle issues, as e.g.\ not any semi-simple
superalgebra is the direct sum of simple superalgebras.
A good review is \cite{Annecy}. The fermionic generators of such superalgebras are
in representations of the bosonic part. If that
`defining representation' of the bosonic algebra in the fermionic generators
is completely
reducible, the algebra is said to be `of classical type'. The others are
`Cartan type superalgebras' $W(n)$, $S(n)$, $\tilde S(n)$ and $H(n)$, which we
will further neglect.
For further reference, I give the list of the real forms of
superalgebras `of classical type'. But, for fixing my notations on
algebras, it might be useful first to recapitulate a similar list for
the bosonic algebras. That is done in table~\ref{tbl:bosRealForms}.
\begin{table}[htb]
  \centering
  \caption{Real forms of simple bosonic Lie algebras. For the exceptional
  algebras, I do not repeat the obvious line that the compact form itself
  is one of the real forms. The second number in the notation for
  these real forms is the number of non-compact $-$ number of compact generators.
  }\label{tbl:bosRealForms}
  \vspace{5mm}
 $ \begin{array}{|l|l|l|} \hline
\mbox{Compact } & \mbox{Real Form} & \mbox{Maximal compact
subalg.}\\ \hline
SU(n) & SU(p,n-p)& SU(p)\times SU(n-p)\times U(1)\\
SU(n) & S\ell (n) & SO(n) \\
SU(2n)& SU^*(2n)    & USp(2n) \\ \hline
SO(n) & SO(p,n-p)& SO(p)\times SO(q) \\
SO(2n)& SO^*(2n)   & U(n) \\ \hline
USp(2n)&Sp(2n)  & U(n) \\
USp(2n)&USp(2p,2n-2p)  & USp(2p)\times USp(2n-2p) \\ \hline
G_{2,-14}& G_{2,2} & SU(2)\times SU(2) \\ \hline
F_{4,-52}& F_{4,-20}& SO(9) \\
F_{4,-52}& F_{4,4}& USp(6)\times SU(2) \\ \hline
E_{6,-78}& E_{6,-26}& F_{4,-52}\\
E_{6,-78}& E_{6,-14}& SO(10)\times SO(2)\\
E_{6,-78}& E_{6,2}& SU(6)\times SU(2)\\
E_{6,-78}& E_{6,6}& USp(8)\\ \hline
E_{7,-133}&E_{7,-25}& E_{6,-78}\times SO(2) \\
E_{7,-133}&E_{7,-5}& SO(12)\times SU(2) \\
E_{7,-133}&E_{7,7}& SU(8) \\ \hline
E_{8,-248}&E_{8,-24}&E_{7,-133}\times SU(2)\\
E_{8,-248}&E_{8,8}&SO(16)\\\hline
\end{array}    $
\end{table}
The conventions which I use for groups is that
$Sp(2n)=Sp(2n,\Rbar)$ (always even entry), and $USp(2m,2n)=U(m,n,\Hbar)$.
$S\ell (n)$ is $S\ell (n,\Rbar)$. Further,
$SU^*(2n)=S\ell (n,\Hbar)$ and $SO^*(2n)=O(n,\Hbar)$.
Note that in these bosonic algebras there are the following
isomorphisms, some of which will be important later\footnote{Note
that the equality sign is not correct for the groups. For the
algebras there are these isomorphism, but it are rather the
covering groups of the orthogonal groups which are mentioned at the
right hand sides.}
\begin{eqnarray}
SO(3)&=& SU(2)=SU^*(2)\,,\nonumber\\
SO(2,1) &=&S\ell (2)=SU(1,1)=Sp(2)\,,  \nonumber\\
SO(4)&=&SU(2)\times SU(2)\,,\qquad SO(2,2)=S\ell (2)\times S\ell (2)\,,
\nonumber\\ &&
SO(3,1)=Sp(2,\Cbar)\,,
\nonumber\\
SO^*(4)&=&SU(1,1)\times SU(2)\,,\nonumber\\
SO(5)&=&USp(4)\,,\qquad SO(4,1)=USp(2,2)\,,\qquad SO(3,2)=Sp(4)\,,\nonumber\\
SO(6)&=&SU(4)\,,\qquad SO(5,1)=SU^*(4)\,,\nonumber\\
&& SO(4,2)=SU(2,2)\,, \qquad
SO(3,3)=S\ell (4)\,,\nonumber\\
SO^*(6)&=&SU(3,1)\,,\nonumber\\
SO^*(8)&=&SO(6,2)\,.
\label{isomBosAlg}
\end{eqnarray}
\par
The similar table for Lie
superalgebras `of classical type', and their real forms \cite{realLieSA} is
given in table~\ref{tbl:LieSA}.
In this table\footnote{The table has been changed after the first version
of this paper, correcting also the table in \cite{Annecy}. These
corrections have been found in discussions with S.~Ferrara.}, `defining
representation' gives the  fermionic generators as a
representation of the bosonic subalgebra. The `number of generators' gives the
numbers of (bosonic,fermionic) generators in the superalgebra. I mention first
the algebra as an algebra over $\Cbar$, and then give different real forms of
these algebras.
With this information, you can reconstruct all properties of these algebras, up
to a few exceptions.
The names which I use for the real forms is for
some algebras different from those in the
mathematical literature \cite{realLieSA}, and chosen such that it is most suggestive of its
bosonic content. There are isomorphisms as $SU(2|1)=OSp(2^*|2,0)$,
and $SU(1,1|1)=S\ell (2|1)=OSp(2|2)$. In the algebra $D(2,1,\alpha)$
the three $S\ell (2)$ factors of the bosonic group in the anticommutator of
the fermionic generators appear with relative weights $1$, $\alpha$
and $-1-\alpha$. The real forms contain respectively
$SO(4)=SU(2)\times SU(2)$, $SO(3,1)=S\ell (2,\Cbar)$ and
$SO(2,2)=S\ell (2)\times S\ell (2)$. In the first and last case
$\alpha$ should be real, while $\alpha= 1 +ia$ with real $a$
for $p=1$. In the
limit $\alpha=1$ one has the isomorphisms $D^p(2,1,1)=OSp(4-p,p|2)$.
\begin{table}[ht]\caption{Lie superalgebras of classical type. For the
real forms of $SU(m|m)$, the one-dimensional subalgebra of the bosonic
algebra is not part of the irreducible algebra. Furthermore, in that case
there are subalgebras obtained from projection of those mentioned here
with only one factor $SU(n)$, $S\ell (n)$, $SU^*(n)$ or $SU(n-p,p)$ as
bosonic algebra.
\label{tbl:LieSA}}
\begin{center}\begin{tabular}{|l|l|l|l|l|}\hline
Name & Range  & Bosonic algebra  & Defining& Number of  \\
     &        &                  & repres. & generators\\ \hline \hline
$SU(m|n)$&$m\geq 2$&$ SU(m)\oplus SU(n)$&$(m,\bar n)\oplus
$&$m^2+n^2-1,$\\
     & $ m\neq n $       & \ \ \ $\oplus U(1)$&\ $(\bar m,n)$ &\ \ \  $2mn  $\\
& $ m= n $&\ \ \ no $U(1)$   & 
&$2(m^2-1),2m^2$\\
     \hline
 \multicolumn{2}{c|}{
\begin{tabular}{|l}
$S\ell(m|n) $       \\
$SU(m-p,p|n-q,q) $   \\
$SU^*(2m|2n) $\\
$S\ell \,{}^\prime(n|n)$
\end{tabular}}
& \multicolumn{3}{l|}{\begin{tabular}{l}
$\begin{array}{l}
S\ell (m)\oplus S\ell (n)  \\
SU (m-p,p)\oplus SU(n-q,q)\\
SU^*(2m)\oplus SU^*(2n)
\end{array}  $
\hspace{-15pt} $\left.
\begin{array}{l}
\oplus SO(1,1)   \\
\oplus U(1)   \\
\oplus SO(1,1)
\end{array} \hspace{-5pt}\right\}$\hspace{-10pt} \begin{tabular}{l}
if  \\
 $m\neq n$
\end{tabular}
\\ $S\ell (n,\Cbar)$\end{tabular}}
\\ \hline\hline
$OSp(m|n)$&$ m\geq 1$ & $SO(m)\oplus Sp(n)$ & $(m,n)$ & $\ft12(m^2-m+$\\
          & $ n=2,4,..$ &                   &         &\ \ \  $n^2+n),mn$\\ \hline
 \multicolumn{2}{c|}{
\begin{tabular}{|l}
\phantom{$SU(m-p,p|n-q,q) $}  \\[-4mm]
$OSp(m-p,p|n)$  \\
$OSp^*(m|n-q,q)$
\end{tabular} }&
 \multicolumn{2}{c}{
\begin{tabular}{l}
\phantom{$SU (m-p,p)\oplus SU(n-q,q)$}\hspace{-5pt}  \\[-4mm]
$ SO(m-p,p)\oplus Sp(n)$ \\
  $ SO^*(m)\oplus USp(n-q,q)$
\end{tabular} }&
\begin{tabular}{l} $n$ even
 \\ $m,n,q$ even
\end{tabular}
\\ \hline\hline
$D(2,1,\alpha)$ & $0<\alpha\leq 1$ & $SO(4)\oplus S\ell (2)$&$(2,2,2)$&9, 8 \\
                \hline
 \multicolumn{2}{c|}{
\begin{tabular}{|l}
\phantom{$SU(m-p,p|n-q,q) $}   \\[-4mm]
$D^p(2,1,\alpha)$
\end{tabular} }&
 \multicolumn{2}{c}{
\begin{tabular}{l}
\phantom{$SU (m-p,p)\oplus SU(n-q,q)$} \\[-4mm]
 $SO(4-p,p) \oplus S\ell (2)$
\end{tabular} }&$p=0,1,2$ \\ \hline\hline
$F(4)$ & & $\overline {SO(7)}\oplus S\ell (2)$ & $(8,2)$& 24, 16 \\ \hline
 \multicolumn{2}{c|}{
\begin{tabular}{|l}
\phantom{$SU(m-p,p|n-q,q) $} \\[-4mm]
$F^p(4)$\\ $F^p(4)$
\end{tabular} }&
 \multicolumn{2}{l}{
\begin{tabular}{l}
\phantom{$SU (m-p,p)\oplus SU(n-q,q)$} \\[-4mm]
$SO(7-p,p)\oplus S\ell (2)$\\ $SO(7-p,p)\oplus SU(2)$
\end{tabular} }&
\begin{tabular}{l}
$p=0,3$    \\ $p=1,2$
\end{tabular}
\\ \hline\hline
$G(3)$ & & $G_2\oplus S\ell (2)$ & $(7,2)$ & 14, 14\\  \hline
 \multicolumn{2}{c|}{
\begin{tabular}{|l}
\phantom{$SU(m-p,p|n-q,q) $} \\[-4mm]
$G_p(3)$
\end{tabular} }&
 \multicolumn{2}{c}{
\begin{tabular}{l}
\phantom{$SU (m-p,p)\oplus SU(n-q,q)$} \\[-4mm]
$G_{2,p}\oplus  S\ell (2)$
\end{tabular}} & $p=-14,2$\\ \hline\hline
$P(m-1)$&$m\geq 3$ & $S\ell (m)$ & $(m\otimes m)$&$m^2-1,m^2$ \\ \hline\hline
$Q(m-1)$&$m\geq 3$ & $SU(m)$ & Adjoint &$m^2-1,m^2-1$ \\ \hline
 \multicolumn{2}{c|}{
\begin{tabular}{|l}
\phantom{$SU(m-p,p|n-q,q) $} \\[-4mm]
$Q(m-1)$ \\
$Q((m-1)^*)$  \\
$UQ(p,m-1-p)$
\end{tabular} }&
 \multicolumn{2}{c}{
\begin{tabular}{l}
\phantom{$SU (m-p,p)\oplus SU(n-q,q)$} \\[-4mm]
$S\ell (m)$\\
 $SU^*(m)$ \\
 $SU(p,m-p)$
\end{tabular}} & \\ \hline\hline
\end{tabular}\end{center} \end{table}
\subsection{Superconformal algebras}\label{ss:scalg}
In this section I review the classification of superconformal algebras of a standard
type, to be explained soon. At the same time, however, also the $adS$
superalgebras of the similar type will be classified, due to the
isomorphism (\ref{confisadS}).
\par
In supersymmetric theories, the conformal symmetry implies the presence of a
second supersymmetry $S$, usually denoted as `special supersymmetry'. Indeed,
the commutator of the special conformal transformations, and the ordinary
supersymmetry $Q$ implies this $S$ due to $[K_\mu,Q]=\Gamma_\mu S$.
\par
The anticommutator $\left\{Q,S\right\}$ generates also an extra
bosonic algebra (sometimes called $R$ symmetry). The whole superalgebra
can be represented in a supermatrix, as e.g.\ (symbolically)
\begin{equation}
\pmatrix{SO(d,2)& Q+S\cr Q-S&R}\ .
\end{equation}
We can consider such superalgebras in general. That is what Nahm \cite{nahm}
did in his classification. The requirements for a superconformal algebra
in $d$ or a super-adS algebra in $d+1$ are:
\begin{enumerate}
\item $SO(d,2)$  should appear as a factored subgroup of the
bosonic part of the superalgebra. For Nahm, this requirement was
motivated by the Coleman--Mandula theorem, but one can demand it also
in order
that the bosonic algebra is the isometry algebra of a space
which has the $adS$ space as a factor.
\item  fermionic generators should
sit in a spinorial representation of that group.
\end{enumerate}

To find the list of superconformal algebras, one has to consider the
 isomorphisms of groups $SO(d,2)$ which are in (\ref{isomBosAlg}).
Then the analysis is straightforward, and the result are algebras
with maximal $d=6$\footnote{Note the particular case of $d=6$, where we use the notation
$OSp(8^*|N)$ for the superconformal algebra. Often, including previous articles of ourselves,
it was written as $OSp(6,2|4)$, not paying attention to the existing real forms.
In fact, in the series $OSp(m-p,p|2n)$ the algebra $Sp(2n)$ is non-compact.
Thus, a compact $R$-symmetry group, $USp(2n)$, is not possible in that series.
The possibility of a compact $R$-symmetry exists
due to the isomorphism $SO^*(8)=SO(6,2)$, such that one can use the next line
of table~\ref{tbl:LieSA}. This thus
works only for the signature (6,2).}. The result\footnote{We mention the superalgebra with
compact $R$-symmetry group.} is given in
Table~\ref{tbl:sca}, except for $d=2$.
For $d=2$ the finite bosonic adS or conformal algebra is $SO(2,2)\approx
SO(2,1)\oplus SO(2,1)$, i.e.\ the sum of two $d=1$ algebras. The super-adS or
superconformal algebra is then the sum of two $d=1$ algebras of
Table~\ref{tbl:sca}.
\begin{table}[t]\caption{Super $adS_{d+1}$ or $conf_d$ algebras.
\label{tbl:sca}}\vspace{0.4cm}
\begin{center}\begin{tabular}{|lllc|}\hline
$d$& superalgebra& $R$ & number of fermionic\\ \hline
$1$&  $OSp(N|2)$ & $O(N)$    & $2N$ \\
   &  $SU(N|1,1)$  &$SU(N)\times U(1)$ for $N\neq 2$ &$ 4N$ \\
   &  $SU(2|1,1)           $    &$SU(2)              $ &$ 8   $\\
   &  $OSp(4^*|2N)         $    &$SU(2)\times USp(2N)$ &$ 8N  $\\
   &  $G(3)                $    &$G_2                $ &$ 14  $\\
   &  $F^0(4)                $    &$SO(7)              $ &$ 16  $\\
   &  $D^0(2,1,\alpha)     $    &$SU(2)\times SU(2)  $ &$  8  $\\  \hline
$3$&$ OSp(N|4)   $ &$ SO(N) $&$ 4N $\\   \hline
$4$&$ SU(2,2|N)  $ &$SU(N)\times U(1)$ for $N\neq 4$&$ 8N$\\
   &$ SU(2,2|4)  $ &$  SU(4)$ & 32\\  \hline
$5$&$ F^2(4)       $ &$ SU(2) $ & 16 \\       \hline
$6$&$ OSp(8^*|N) $ &$  USp(N)\ \ (N $ even)& $8N$ \\
\hline
\end{tabular}
\end{center}
\end{table}
Notice that these are the finite part of infinite dimensional superconformal
algebras in 2 dimensions. For a classification of the infinite
superconformal algebras, see \cite{classinf}. One may
also relax the condition that the bosonic algebra contains the algebra
$SO(d,2)$ as a factored subgroup of the whole bosonic algebra, and suffice with
having it as some subgroup. Then the other bosonic symmetries are not
necessarily scalars and the Coleman--Mandula theorem is violated. However, this
may still be relevant where branes are present and has been used e.g.\ in
\cite{JWvHAVP} to propose the $OSp(1|32)$ as super $adS_{11}$ or $conf_{10}$.
In that case one has (\ref{algPoind11cc})
This algebra is now known as the $M$-theory algebra \cite{Mfromsa}.

To recognize it as $OSp(1|32)$, one can write it as
\begin{eqnarray}\label{commMQ}
&&\left\{ Q_\alpha ,Q_\beta \right\}=M_{\alpha \beta }\,,\qquad
  \left[M_{\alpha \beta },Q_\gamma \right]= Q_{(\alpha }{\cal C}_{\beta )\gamma }\nonumber\\
&&  \left[ M_{\alpha \beta },M_{\gamma \delta }\right] = {\cal C}_{\alpha (\gamma }
M_{\delta )\beta }+ {\cal C}_{\beta (\gamma }M_{\delta )\alpha }\,.
\end{eqnarray}
\textit{Exercise:} Obtain the commutators in (\ref{superadSd11}) and the
Lorentz algebra from
this general rule, defining first $M_{\alpha \beta }$ as the right
hand side of the anticommutator of the supersymmetries. \QED
In this way it is clear how all the bosonic generators form $Sp(32)$
where the antisymmetric metric in that algebra is $ {\cal C}_{\alpha \beta
}$.

These superconformal algebras are used in adS/CFT correspondence,
or other applications in M-theory. I have been first interested in
them for the construction of general matter couplings in super-\Poin\
theories. This goes along the lines explained at the end of
section~\ref{ss:localconf}.
For reviews on this method in 4 dimensions, see
\cite{revN2,Karpacz2}. For 6 dimensions it has be applied for
 (1,0) supersymmetry in
\cite{01d6,d6SC} and for (2,0) supersymmetry in \cite{20d6}.

\medskip
\section*{Acknowledgments.}

\noindent
I am grateful to the organizers of the school in
C\v{a}lim\v{a}ne\c{s}ti, especially to Radu Constantinescu, for this very
nice experience.
The lectures already circulated for some time in preliminary form, and
have been improved thanks to useful remarks of Pieter-Jan De Smet,
Joeri Kenes, and Mauricio Werneck de Oliveira. Furthermore, I thank Eric
Bergshoeff, Piet Claus, Sergio Ferrara, Joaquim Gomis, Renata Kallosh
 and Walter Troost for useful discussions.
This work was
supported in part by the European Commission TMR programme
ERBFMRX-CT96-0045.
\newpage

\appendix
\section{Conventions}  \label{app:conv}
I use the metric signature $(-+\ldots +)$.
If you prefer the opposite,
insert a minus sign for every upper index which you see, or for an
explicit metric $\eta_{ab}$ or $g_{\mu\nu}$. The Gamma matrices $\Gamma
_a$ should then be multiplied by an $\rmi$ to have this change of
signature.

The curved indices are  denoted by
$\mu, \nu, \ldots = 0,\ldots , d-1$ and the flat ones by $a, b, ...$.
(Anti)symmetrization is done with weight one: $A_{[ab]}=\ft12 \left(
A_{ab}-A_{ba}\right) $ and $A_{(ab)}=\ft12 \left(
A_{ab}+A_{ba}\right) $.

The anticommuting Levi--Civita tensor is taken to be
\begin{equation}
\varepsilon_{12\ldots d}=1\,,\qquad \varepsilon^{12\ldots d}=(-)^t\,,
\end{equation}
where $t$ is the number of timelike directions. If we use the index value $0$
for Minkowski space, then $\varepsilon_{01\ldots (d-1)}=1$. The contraction
identity for these tensors is ($p+n=d$)
\begin{equation}
\varepsilon_{a_1\ldots a_nb_1\ldots b_p} \varepsilon^{a_1\ldots a_nc_1\ldots c_p}=
(-)^t\, p!\, n!\, \delta_{[b_1}^{[c_1}\ldots \delta_{b_p]}^{c_p]}\,.
\end{equation}
For the local case, we can still define constant tensors
\begin{equation}
\varepsilon_{\mu_1\ldots \mu_d}=e^{-1}e_{\mu_1}^{a_1}\ldots e_{\mu_d}^{a_d}
\varepsilon_{a_1\ldots a_ d}   \,,\qquad
\varepsilon^{\mu_1\ldots \mu_d}=e e^{\mu_1}_{a_1}\ldots e^{\mu_d}_{a_d}
\varepsilon^{a_1\ldots a_ d}   \,.\label{LeviCivLocal}
\end{equation}
They are thus \textit{not} obtained from each other by raising or lowering indices with
the metric.

\noindent \textit{Exercises.}
\begin{enumerate}
  \item Show that the tensors in (\ref{LeviCivLocal})
are indeed constants, i.e.\ that arbitrary variations of the vierbein
cancel in the full expression. You will have to use the so-called
`Schouten identities', which means that antisymmetrizing in more
indices than the range of the indices, gives zero.
  \item If one defines in even dimensions $d=2n$ the dual of an
  $n$-tensor as
\begin{equation}
  \tilde F_{a_1\ldots a_n}= (\rmi)^{d/2+t} \frac{1}{n!}
  \varepsilon _{a_1\ldots a_d}F^{a_d\ldots a_{n+1}}\,,
\label{dualF}
\end{equation}
check that
\begin{equation}
  \tilde {\tilde F}_{a_1\ldots a_n} =F_{a_1\ldots a_n}\,,\qquad
\Gamma _{a_1\ldots a_n}=\Gamma _*\tilde \Gamma _{a_1\ldots a_n}\,.
\label{dualrelations}
\end{equation}
\end{enumerate}

\end{document}